\documentclass{aa}

\usepackage{graphics}
\usepackage{multirow}

\begin{document}

%
%
\title {
Chandra LETGS Observation of the Active Binary Algol}
\author{
        J.-U. Ness\inst{1}
        \and
        J.H.M.M. Schmitt\inst{1}
        \and
        V. Burwitz\inst{2}
        \and
        R. Mewe\inst{3}
        \and
        P. Predehl\inst{2}
        }

\institute{
 Universit\"at Hamburg, Gojenbergsweg 112, D-21029 Hamburg, Germany
 \and
  Max-Planck-Institut f\"ur Extraterrestrische Physik (MPE), Postfach 1603,
  D-85740 Garching, Germany
  \and
 Space Research Organization Netherlands (SRON),
 Sorbonnelaan 2, 3584 CA Utrecht, The Netherlands
}

\authorrunning{J.U. Ness, J.H.M.M. Schmitt, et al.}
\titlerunning{Chandra observations of Algol}

\offprints{J.-U.\ Ness}
\mail{jness@hs.uni-hamburg.de}
\date{received \today; accepted $\ldots$}

%
\abstract{
A high-resolution spectrum obtained with the low-energy transmission grating
onboard the Chandra observatory is presented and analyzed. Our analysis
indicates very hot plasma with temperatures up to $T\approx 15-20$\,MK from the
continuum and from ratios of hydrogen-like and helium-like ions of Si, Mg, and
Ne. In addition lower temperature material is present since O\,{\sc vii} and
N\,{\sc vi} are detected. Two methods for
density diagnostics are applied. The He-like triplets from N\,{\sc vii} to
Si\,{\sc xiii} are used and densities around $10^{11}$\,cm$^{-3}$ are found for
the low temperature ions. Taking the UV radiation field from the B star
companion into account, we find that the low-Z ions can be affected by the
radiation field quite strongly, such that densities of $3\times
10^{10}$\,cm$^{-3}$ are
also possible, but only assuming that the emitting plasma is immersed in the
radiation field. For the high temperature He-like ions only low density limits
are found. Using ratios of Fe\,{\sc xxi} lines produced at similar
temperatures are sensitive to lower densities but again yield only low
density limits. We thus conclude that the hot plasma has densities below
$10^{12}$\,cm$^{-3}$. Assuming a constant pressure corona we show that the
characteristic loop sizes must be small compared to the stellar radius and that
filling factors below 0.1 are unlikely.
}
\maketitle
%

\section{Introduction}

Stellar coronae cannot be spatially resolved, yet they
are thought to be highly structured just like the solar corona, whose 
X-ray emission comes almost exclusively from hot plasma confined in
magnetic loops. So far the only way to infer
structural information in such unresolved stellar point sources has been via
eclipse studies in suitably chosen binary systems.
Observations of the X-ray light curve can yield information
on the location of the X-ray emission (\cite{pres95} 1995),
although the eclipse mapping
reconstruction problem is highly under-determined; after all, one is trying 
to reconstruct a three-dimensional intensity distribution from a 
one-dimensional light curve. Among a variety of problems discussed
in detail by \cite{schmitt98} (1998), a specific difficulty arises 
from the fact
that in most eclipsing systems both components are known or likely
to be X-ray emitters. Obviously the reconstruction problem is easier
to solve in those cases where one of the binary components is X-ray dark.
At X-ray wavelengths
only two such systems have been studied so far, the eclipsing
binary systems $\alpha$ CrB (cf. \cite{sk93} 1993) and $\beta$ Per 
(= Algol; \cite{oord89} 1989).

The Algol system actually consists of three components, a close eclipsing
binary (containing a B8 main sequence star and a K2IV subgiant) and a 
more distant F-type star, which is not of interest for 
our purposes. The stellar parameters of
the two stars of the eclipsing binary system 
(inclination angle is i=81$^\circ$)
are listed in Tab.~\ref{star_prop}.
Algol is one of the brightest coronal X-ray emitters in the soft X-ray band
and has been observed with essentially all X-ray satellites flown so far.
Particular interest in Algol's X-ray emission
arises from the fact that no magnetic dynamos and magnetic activity phenomena
should occur on stars of spectral type B8, since such stars are fully
radiative and thus the primary component of Algol should be X-ray dark. In
consequence, all of Algol's X-ray emission is believed to originate from the
cool secondary, which is rapidly rotating because it is tidally locked with the
primary on the orbital time scale (2.8 days). We note in passing, however, that
there is -- in contrast to the totally eclipsing system $\alpha$ CrB (cf.
\cite{sk93} 1993) -- no observational proof for this assumption.
Nevertheless, X-ray eclipses at secondary optical minimum are expected, yet
not all observations of Algol at secondary minimum yield evidence for
such eclipses. For example, a long observation of Algol with the
EXOSAT satellite (\cite{white86} 1986) centered on secondary optical minimum
showed no indication for any eclipse,
suggesting the interpretation of a corona with a scale height of more 
than a stellar radius or a somewhat peculiar configuration of the corona
at the time of observation. On the other hand, a long ROSAT PSPC observation
(\cite{ott96} 1996) did show evidence for a partial eclipse of the quiescent
X-ray emission, demonstrating that a significant fraction of the quiescent
X-ray emission is emitted within a stellar radius. A BeppoSAX observation
(\cite{schmitt99} 1999, \cite{fav99} 1999) of Algol showed the total eclipse of
a long-duration flare, and a sequence of four ASCA observations of Algol at
secondary eclipse showed evidence for both eclipses and absence of eclipses at
different occasions.

Another method to provide information on structure in spatially unresolved data
consists of spectroscopic measurements of density.
If the density measurements are combined with the measurement of the
volume emission measure EM, an estimate of the emitting plasma volume
can be obtained; in an eclipsing binary these volumes will be subject to
additional light curve constraints.
With the high-resolution spectrometers onboard {\it Chandra} it is
possible to carry out high-resolution X-ray spectroscopy for a wide range of
coronal X-ray sources. We have obtained
a {\it Chandra} high-resolution X-ray spectrum of Algol,
which allows us to combine the information derived from X-ray light curves
and X-ray spectroscopy. We will specifically discuss the Algol spectra
obtained with the Low Energy Transmission Grating Spectrometer (LETGS).\\

\begin{table}
\caption[ ]{\label{star_prop}Properties of Algol A and B:
mass $M$, radius $R$, effective temperature $T_{\rm eff}$, log\,g and
spectral type are taken from \cite{rich93} (1993) and references therein.}
\renewcommand{\arraystretch}{1.2}
\begin{tabular}{r r r}
\hline
&Algol A & Algol B \\
\hline
d/pc        &  \multicolumn{2}{c}{28}\\
$M/M_\odot$ &     $3.7\pm0.3$  &  $0.81\pm0.05$\\
$R/R_\odot$ &     $2.9\pm0.04$ &  $3.5\pm0.1$  \\
$T_{\rm eff}$/K&  $13000\pm500$ & $4500\pm300$ \\
log\,g       &     $4.08$	& $3.2$	 \\
Spectr. type & B8V       &  K2IV    \\
\hline
\end{tabular}
\begin{flushleft}
\renewcommand{\arraystretch}{1}
\end{flushleft}
\end{table}

\section{Instrument description and Observation}

\begin{figure*}
  \resizebox{\hsize}{!}{\includegraphics{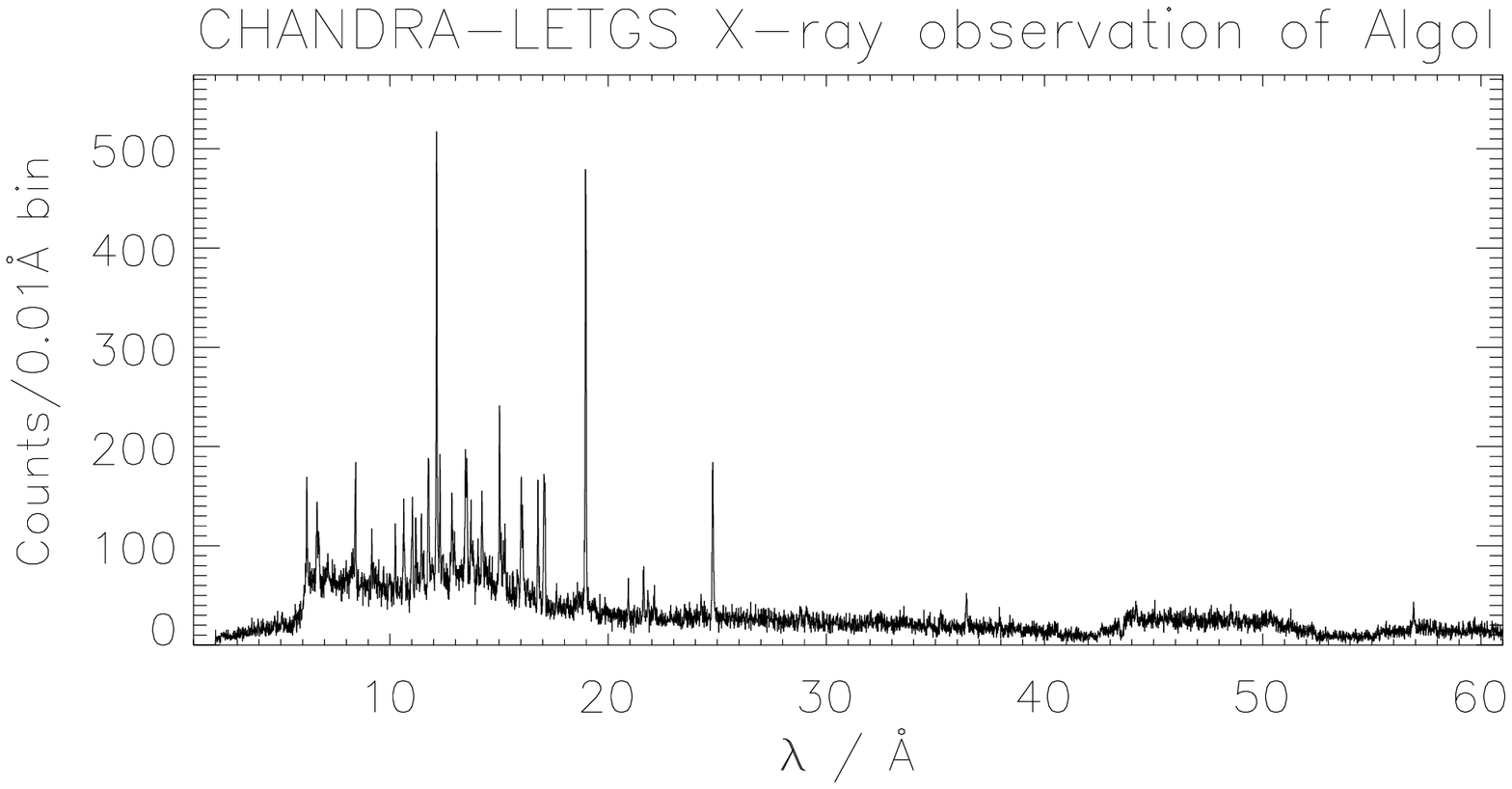}}
  \resizebox{\hsize}{!}{\includegraphics{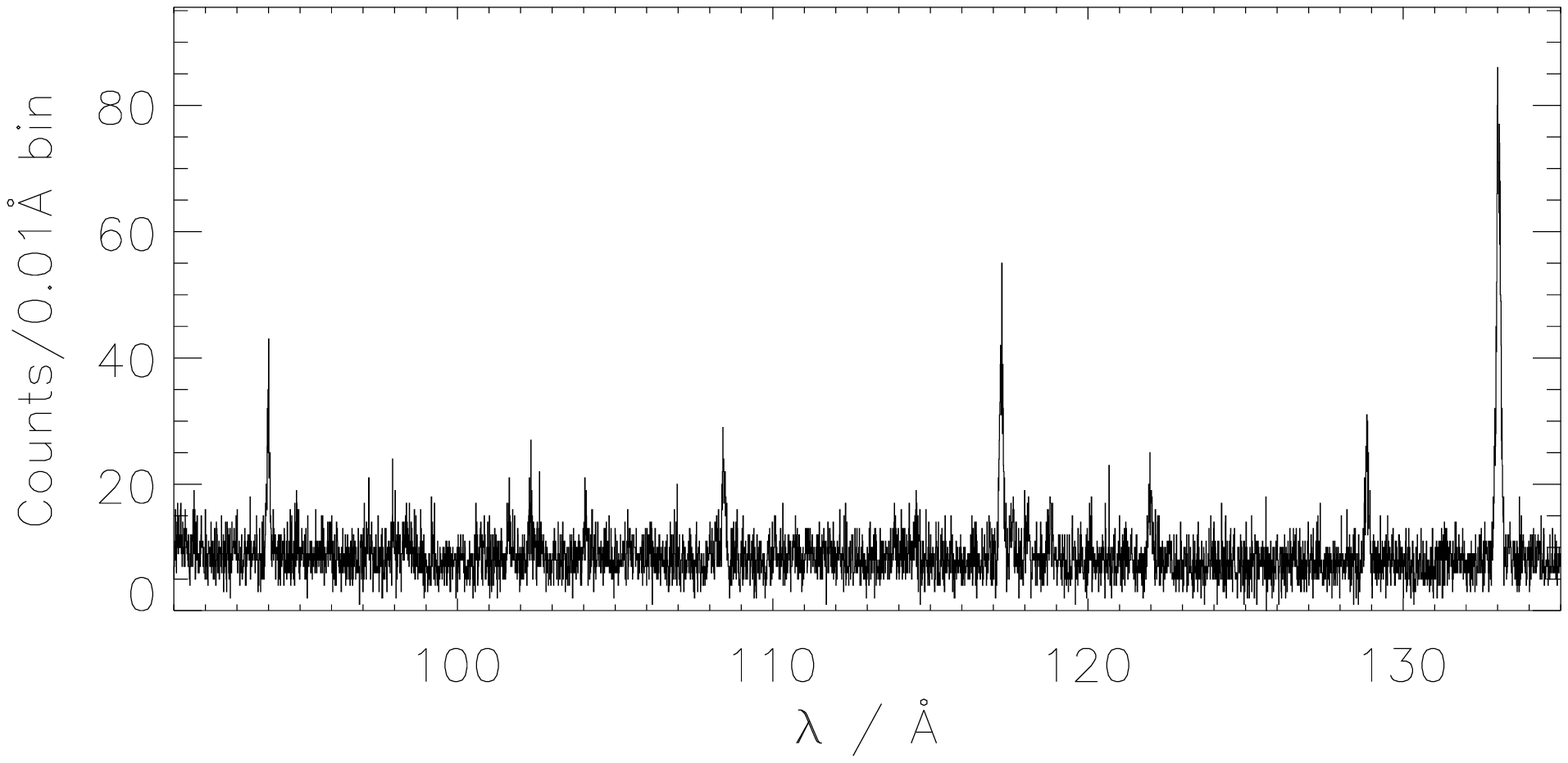}}
 \caption[]{{\bf Top}: LETGS spectrum of Algol in the range 1-40\,\AA. Clearly
visible is the strong continuum and many prominent emission lines from Ne, Si,
Fe, Mg, O, and N. The Ly$_\alpha$ lines from Mg\,{\sc xii} (8.42\AA),
Ne\,{\sc x} (12.14\AA), O\,{\sc viii} (18.97\AA) and N\,{\sc vi} (24.78\AA) can
be recognized.\\
{\bf Bottom}: Wavelength range from 90-135\,\AA. Visible are the lines from
Fe\,{\sc xviii} at 94\,\AA\, from Fe\,{\sc xxi} at 117.5\,\AA\ and 128.7\,\AA, and
from Fe\,{\sc xxiii}/Fe\,{\sc xx} at 133\AA.}
 \label{sp_all}
\end{figure*}

The LETGS is a diffraction grating spectrometer covering 
the wavelength range between 5 -- 175\,\AA\ (0.07 -- 2.5~keV) with a resolution
$\lambda /\Delta \lambda \sim 2000 $ at the long wavelength end of the band
pass; typical instrumental line widths are of the order 0.06\,\AA\ (FWHM) (cf.,
\cite{ness01a} 2001a). A detailed description of the LETGS instrument 
is presented by \cite{predehl97} (1997). We note in passing that the LETGS uses
a microchannel plate detector (HRC-S) placed behind the 
transmission grating without any significant intrinsic energy resolution. Thus
in contrast to CCD based detectors the energy information for individual
counting events is solely contained in the events' spatial location.\\
Accounting for the instrumental line widths the symmetry of the grating
is sufficient to co-add both sides of the spectrum in order to obtain a better
SN ratio (\cite{ness01a} 2001a). The thus obtained spectrum is shown in
Fig.~\ref{sp_all}. A rich line spectrum with lines from Fe, Si, Ne, O, and N
can be recognized between
6 and 30\,\AA\ as well as many Fe lines above 90\,\AA. Strong continuum
emission is also apparent almost over the whole observed band pass; we note in 
passing that the spectrum shown in Fig.~\ref{sp_all} has not been corrected 
for effective areas.

\section{X-ray fluxes and light curve}
\label{lc}

\begin{figure}
  \resizebox{\hsize}{!}{\includegraphics{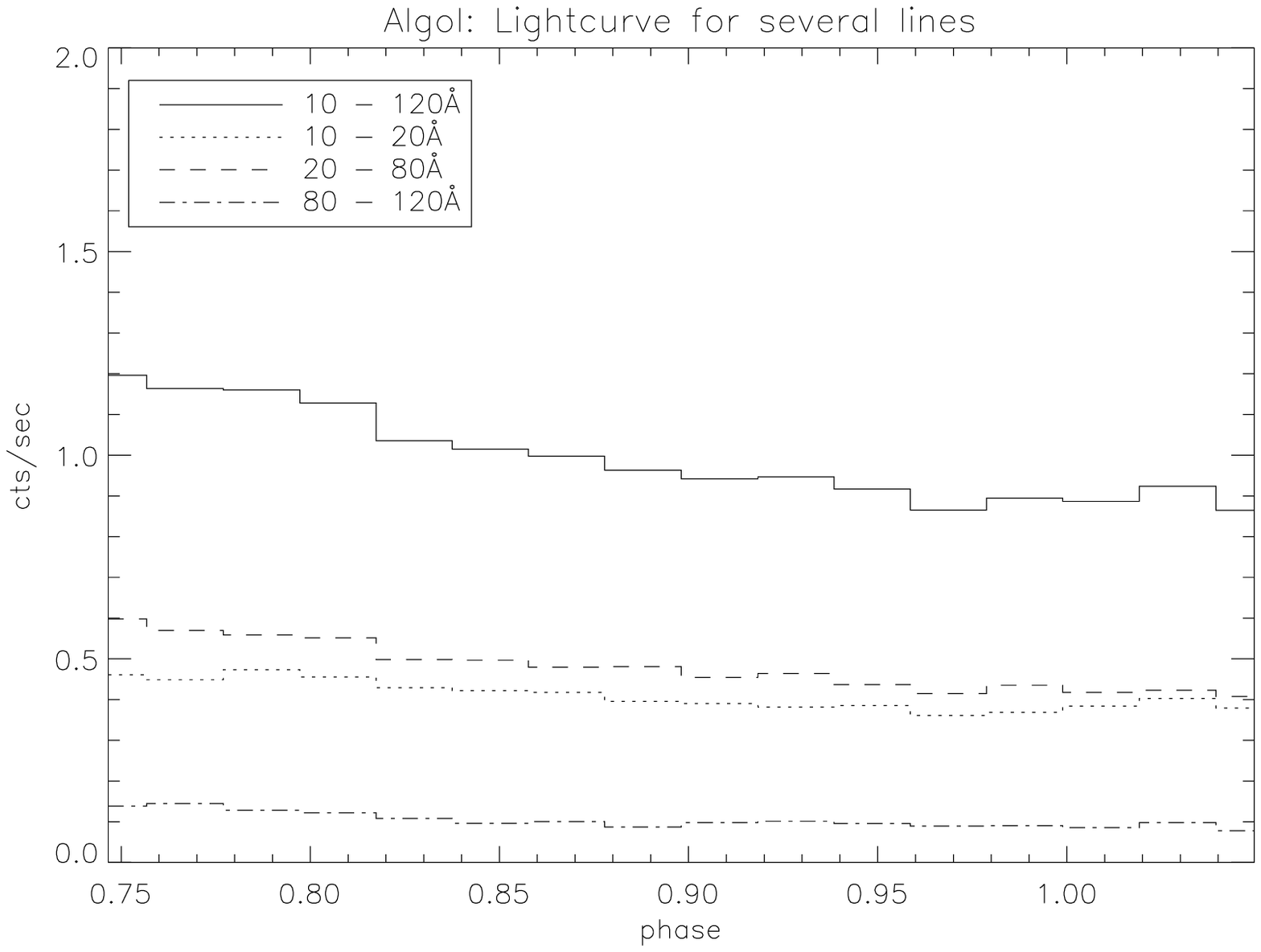}}
 \caption{Light curve of Algol in a hard, medium and a soft energy band.}
 \label{lightcurve1}
\end{figure}

\begin{figure}
  \resizebox{\hsize}{!}{\includegraphics{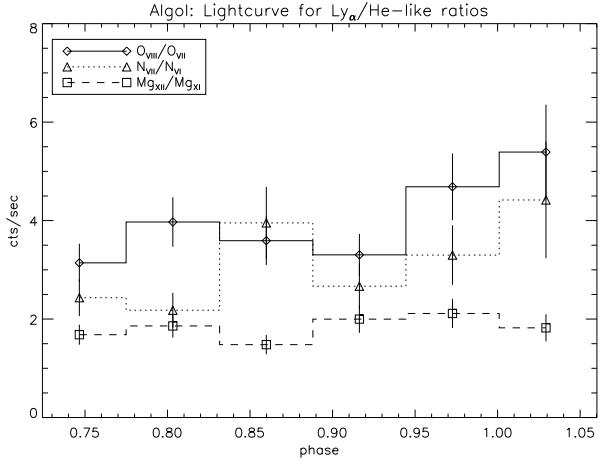}}
 \caption{Light curve of Algol in three different energy bands (top)
and development of the temperature dependent ratios of Ly$_\alpha$/He-like
resonance line for Oxygen, Nitrogen and Magnesium.}
 \label{lightcurve2}
\end{figure}

Algol was observed with the above described instrumental setup between
March, 12, 2000, 18:36 and March, 13, 2000, 17:13. The total on-time
was 81.41\,ksec, almost identical to the actual exposure time.
During this time 106181 source counts were collected on the negative
side and 101469 counts on the positive side. The high spectral resolution 
of the LETGS allows the computation of incident photon and energy fluxes
without the need of any plasma emission model. In particular, 
since the LETGS wavelength range covers both the ROSAT and the Einstein
wavelength ranges we can directly calculate fluxes corresponding 
to the respective band passes of these instruments without the need of 
any model. Using only bins with $A_{\rm eff}>0.1$\,cm$^2$ and the distance d
from Tab.~\ref{star_prop} we compute a total X-ray luminosity of
$1.4\,10^{31}$\,erg/s. Restricting the wavelength range to the nominal ROSAT
wavelength range (6.2-124\,\AA), we find $1.1\,10^{31}$\,erg/s, within the
Einstein band pass we find $1.0\,10^{31}$\,erg/s (2.8-62\,\AA). These numbers
can be compared with earlier measurements with these instruments.
\cite{bergh96} (1996) report an X-ray flux of L$_X=0.7\,10^{31}$\,erg/s measured
with ROSAT. This agrees with our {\it Chandra} measurement to within 30\% so
that significant long term variability can be excluded. \cite{ott96} (1996)
report an X-ray luminosity of L$_X=20\,10^{31}$\,erg/s during a flare, while
their quiescent emission is consistent with the values reported by
\cite{bergh96} (1996). Our measurement is therefore well within the range of
luminosities found in earlier observations.\\
In order to compute the ephemeris of Algol, we used the expression
$JD_{prim}$ = 2445739.003 + 2.8673285 $\times$ E (\cite{kim89} 1989; 
E being an integer) for the times of primary minimum.
Our {\it Chandra} observation covers the phases 0.74 to 1.06, i.e., outside
optical secondary minimum.
In Fig.~\ref{lightcurve1} we show the background-subtracted X-ray light curve
of the LETGS data (in the ranges 10 - 120\,\AA, 10 - 20\,\AA, 20 - 80\,\AA, and
80 - 120\,\AA). As is clear from Fig.~\ref{lightcurve1}, the light curve
shows a more or less continuous decrease in intensity throughout the
{\it Chandra} observations by a factor of 1.38 (10 - 120\,\AA), 1.22 (10 -
20\,\AA), 1.47 (20 - 80\,\AA), and 1.77 (80 - 120\,\AA). Phasing of the data 
suggests the existence of a primary minimum in X-rays, when the late-type
star is located in front of the early-type star, but from our 
discussion above this appears highly unlikely. Since the Algol system is
known to be able to produce giant flares, a far more plausible
interpretation would be to interpret the X-ray light curve as the ``tail'' of a
long-duration flare, possibly similar to the one observed by \cite{schmitt99}
(1999) with BeppoSAX. From this assumption, however, we would expect
the radiation to become softer in time and to detect a cooling of the plasma.
From Fig.~\ref{lightcurve1} no evidence for softening can be deduced, rather
the radiation becomes even harder. From the temperature dependent line ratios
of the resonance lines of the H-like and He-like ions plotted in
Fig.~\ref{lightcurve2} again no evidence for cooling is apparent; the plasma
might even become hotter with time, but at a very small rate. For oxygen the
ratio raises from 3.6 $\pm$ 0.27 to 4.2 $\pm$ 0.35, for nitrogen from 2.7 $\pm$
0.25 to 3.3 $\pm$ 0.37 and for magnesium from 1.7 $\pm$ 0.12 to 2.0 $\pm$ 0.16.
Thus we find no indication for the ''tail'' of a long duration flare from this
spectral analysis. Other plausible scenarios can be thought of as, e.g., the
time-evolution of one or more of the coronal active regions or rotational
modulation.\\



\section{Data analysis}

The data extraction from the HRC-S and analysis of the spectra presented
in this paper are identical to the methods described by \cite{ness01a} (2001a).
Specifically the spectra are extracted along the spectral trace without
any pulse height correction scheme, the background is taken from nearby
adjacent regions on the microchannel plate. The two dispersion directions are
co-added, but the individual dispersed spectra can still be used to check for
inconsistencies in the co-added spectrum. The thus obtained spectrum is shown
in Fig.~\ref{sp_all}. 

\subsection{Analysis of the continuum}
\label{cont}

\begin{figure}[!ht]
 \resizebox{\hsize}{!}{\includegraphics{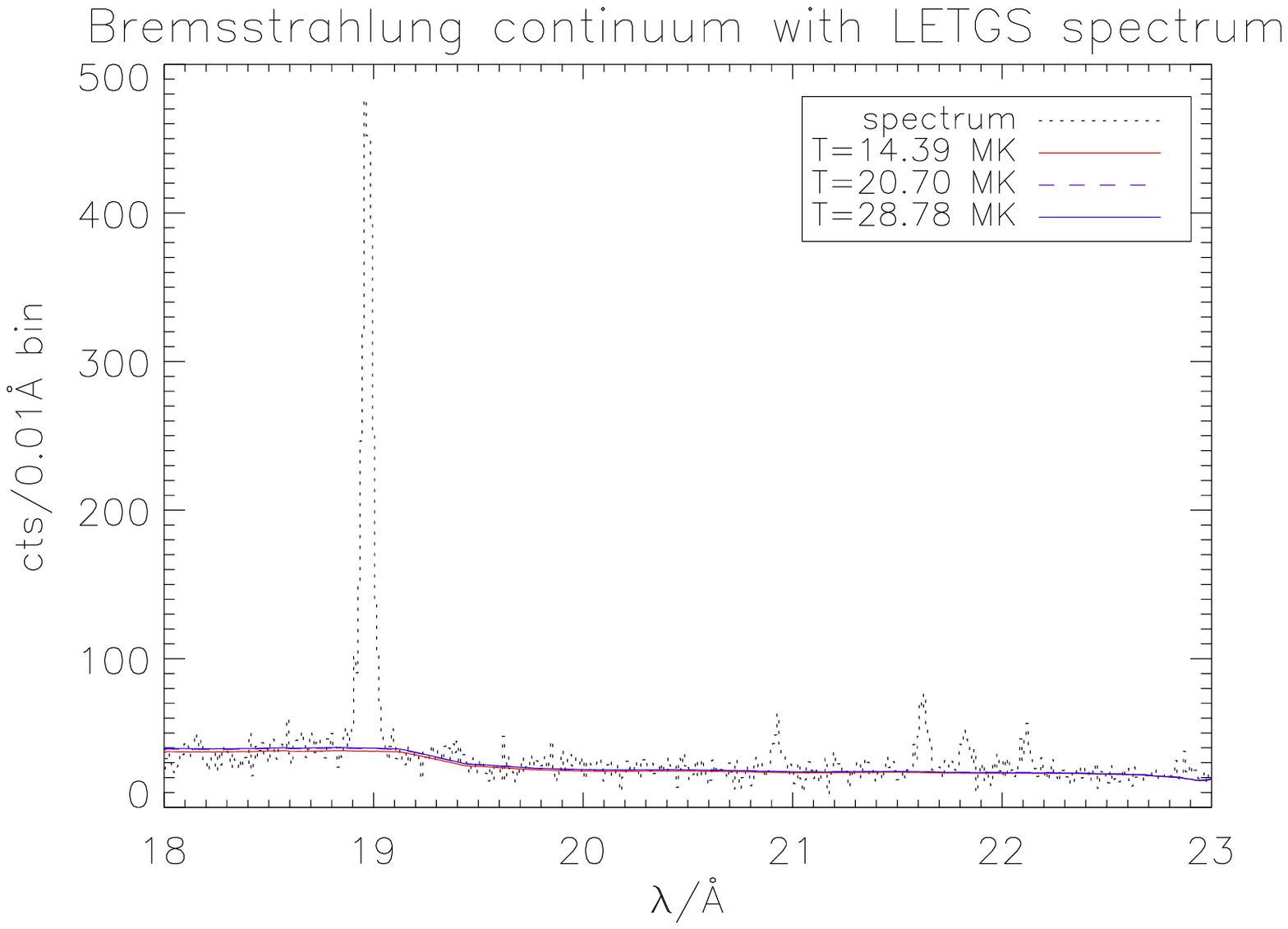}}
 \resizebox{\hsize}{!}{\includegraphics{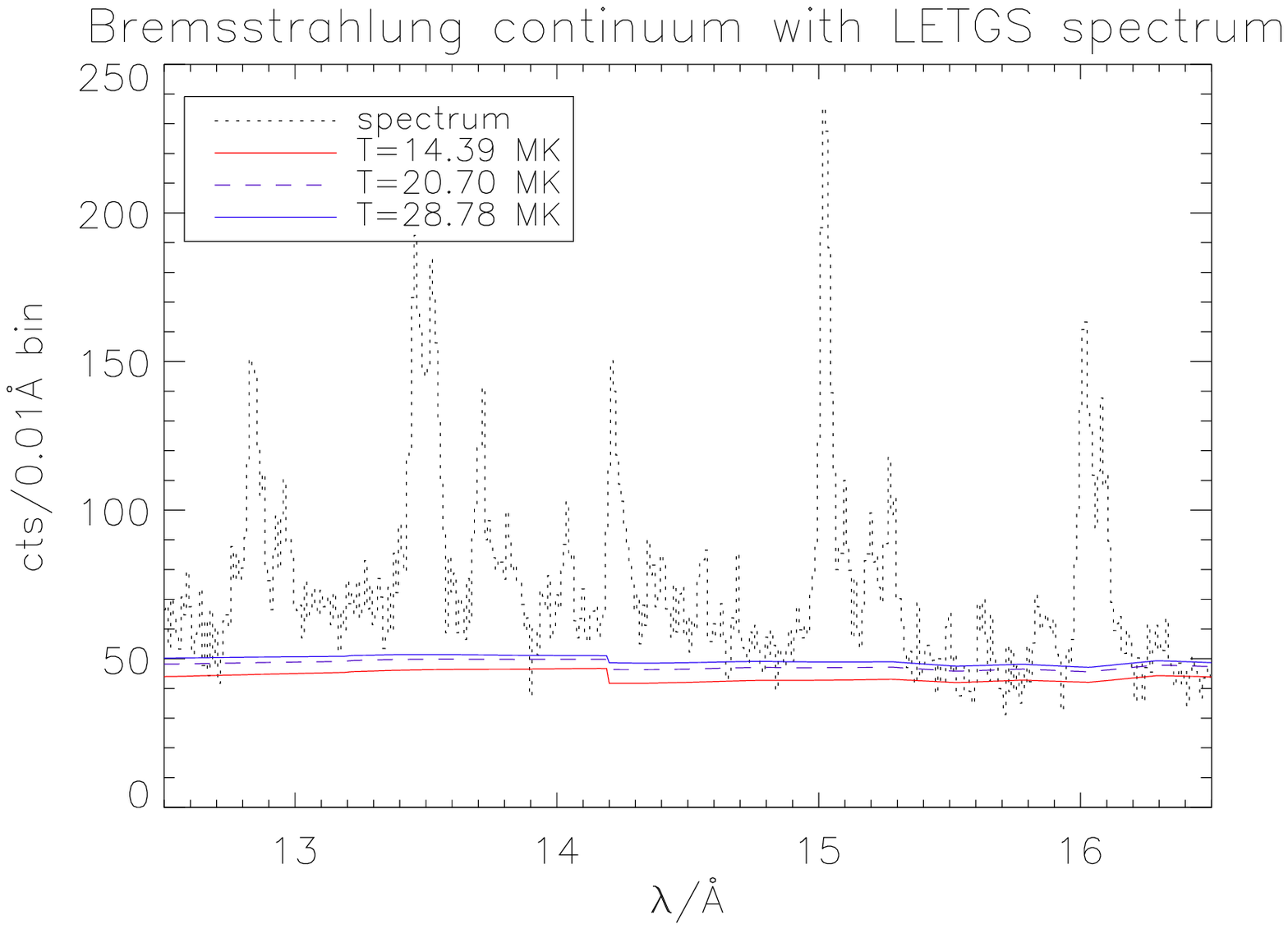}}
 \resizebox{\hsize}{!}{\includegraphics{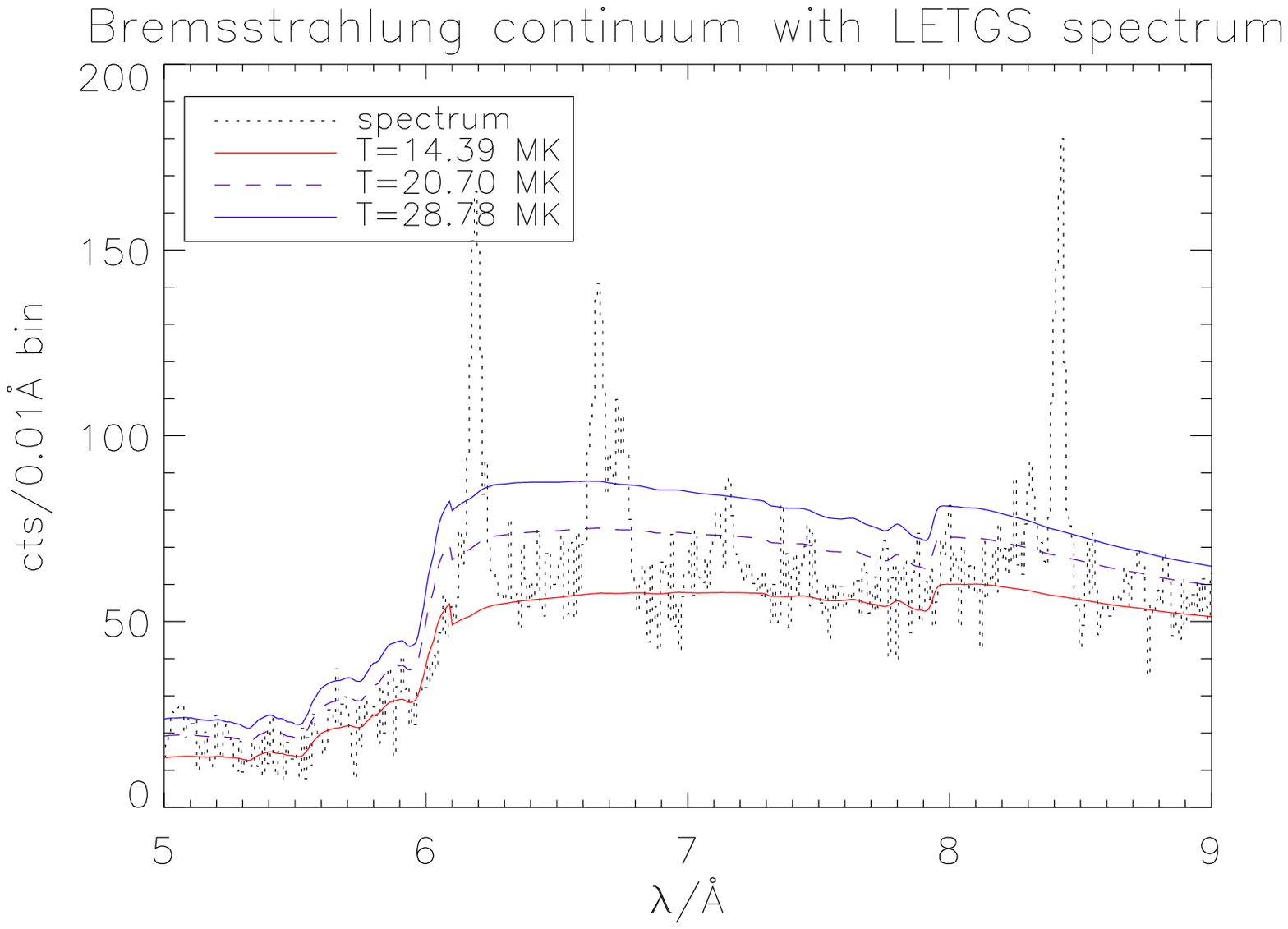}}
 \caption{Modeled bremsstrahlung spectrum with $T=14$, 21, and 29\,MK and
$EM=68\times 10^{52}$\,cm$^{-3}$ in comparison with the measured spectrum.
The model is converted to LETGS counts by use of the effective areas from
\cite{dpease00} (Oct. 2000). Top: very little sensitivity to the temperature,
middle: attempt to model overlaps with higher temperatures, bottom: strong
dependence of temperature suggesting 15\,MK the only temperature consistent with
all wavelength ranges.}
 \label{contspec}
\end{figure}

The total spectrum in Fig.~\ref{sp_all} shows two components, a multitude of
emission lines and a significant continuum. The shape of the continuum suggests
thermal bremsstrahlung emission as the dominant continuum emission
process. This assumption is supported by the high temperatures
measured for Si, Mg, and Fe (cf. Tab.~\ref{ftemp}). We therefore use the formula
\begin{eqnarray}
\label{brems1}
\frac{dW}{dV\ dt\ d\lambda}&=&2.051\times10^{-19}n_e^2 T^{-1/2}\frac{G(\lambda,T)}{\lambda^2}e^{-143.9/(T\lambda)}\nonumber\\
&&\mbox{\footnotesize[erg\,cm$^{-3}$\,s$^{-1}$\,\AA$^{-1}$]}
\end{eqnarray}
(\cite{mewe86} 1986) for the specific volume emissivity $\frac{dW}{dV\ dt\
d\lambda}$ to model the continuum. $T$ is the plasma temperature in K and $c$
the speed of light in cm/s.
The Gaunt factor $G(\lambda,T) = G_{\rm ff}(\lambda,T) + G_{\rm bf}(\lambda,T)$
is modeled as the sum of the free-free Gaunt factor
\begin{eqnarray}
\label{gauntff}
\log\,G_{\rm ff}(\lambda,T)&=&0.355\lambda^{-0.06} \log\lambda\\
&&+0.3\lambda^{-0.066} \log\,\left(\frac{T}{10^8}\right)+0.0043\nonumber
\end{eqnarray}
(\cite{mewe86} 1986) and the bound-free Gaunt factor
\begin{equation}
\label{gauntbf}
G_{\rm bf}(\lambda,T)=aT^b e^{cT^d}
\end{equation}
with the parameters $a$, $b$, $c$, and $d$ adopted from \cite{mewe86} (1986).
Introducing the variable $x=1/\lambda$ we derive from
Eq.~\ref{brems1} an expression for the total number $dW$
of recorded photons divided by effective areas in each wavelength bin:
\begin{equation}
\label{brems2}
dW=B*G(\lambda,143.9/\alpha)/\sqrt{143.9/\alpha}*x^2e^{-\alpha x}
\end{equation}
with $B=2.05\times10^{-19}EM/(4\pi d^2)*\Delta t*\Delta \lambda$, 
$\alpha=143.9/T$, and
$x=1/\lambda$. The plasma temperature can be derived from $\alpha$ and the
emission measure $EM=n^2V$ from $B$ while
$\Delta \lambda$=0.01\,\AA\ (the binsize of the spectrum), and the exposure time
$\Delta t$=81.41\,ksec. Best fit parameters $B$ and $\alpha$ are obtained with a
$\chi^2$ fit. In order to minimize the effects from the emission lines, we
actually use the inverse spectrum to be compared with $1/dW(B,\alpha)$.
In each iteration step we calculate
the temperature from $\alpha$ with $T(\alpha)=143.9/\alpha$ and the Gaunt factor
in each wavelength bin from Eqns.~\ref{gauntff} and \ref{gauntbf}.
From our best fit results we find $T=20.7$\,MK and $EM=67.5\times
10^{52}$\,cm$^{-3}$. This temperature is high, we note, however, that the
sensitivity to the temperature is quite different in the different wavelength
ranges. In Fig.~\ref{contspec} three such wavelength ranges are shown for
three different choices of temperature and the same emission measure
$EM=68\times 10^{52}$\,cm$^{-3}$.
Because of the flatness of bremsstrahlung spectra
the choice of the temperature is not important above 18\,\AA, but the
position of the cutoff does depend sensitively on temperature.
The number of emission lines between 13 and
16\,\AA\ is quite high leading to line blends; the best fit bremsstrahlung 
continuum is therefore low (cf., middle panel of Fig.~\ref{contspec}). 
With the bottom panel we wish
to show that the assumption of $T=14.4$\,MK provides a realistic description
for all wavelength ranges and in particular for the region near the thermal 
cutoff of the X-ray spectrum. In this case 79\% of the X-ray luminosity $L_X$
belongs to the
continuum, thus $L_{X, cont}=1.1\times 10^{31}$\,erg/sec. As a cross check we
calculate the emission measure from the total luminosity and a mean radiative
power loss due to bremsstrahlung of $1.5\times 10^{-23}$\,erg\,cm$^3$/sec, and
obtain $EM=93\times 10^{52}$\,cm$^{-3}$. From this exercise we note that the
continuum can well be modeled with a bremsstrahlung spectrum, and our 
{\it Chandra} data are
consistent with an emission measure $EM=68\times 10^{52}$\,cm$^{-3}$ and a
(peak) temperature at 15\,MK.

\subsection{Extracted spectra and measured line ratios}

\begin{figure}[!ht]
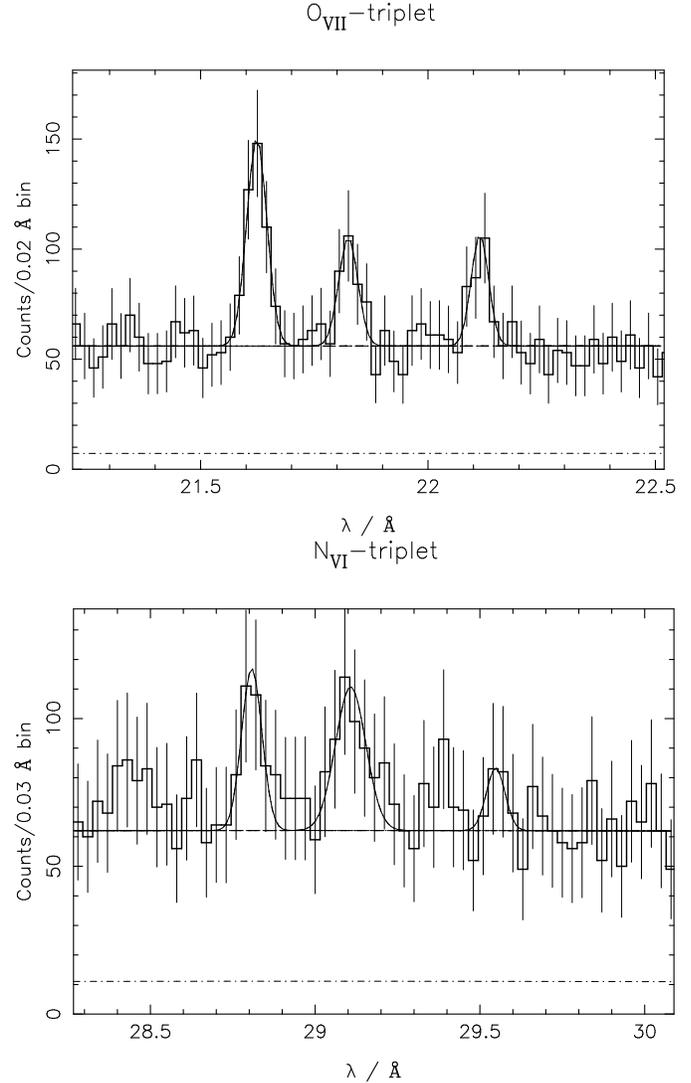

 \resizebox{\hsize}{!}{\rotatebox{270}{\includegraphics{MS2207f8}}}
 \resizebox{\hsize}{!}{\rotatebox{270}{\includegraphics{MS2207f9}}}
 \caption{(a,b) Spectrum (bold line) and best fit (thin solid line) for the triplets
O\,{\sc vii} (a) and N\,{\sc vi} (b)
for Algol. The dashed-dotted line represents the instrumental background.}
 \label{spectra_on}
\end{figure}
\begin{figure}[!ht]
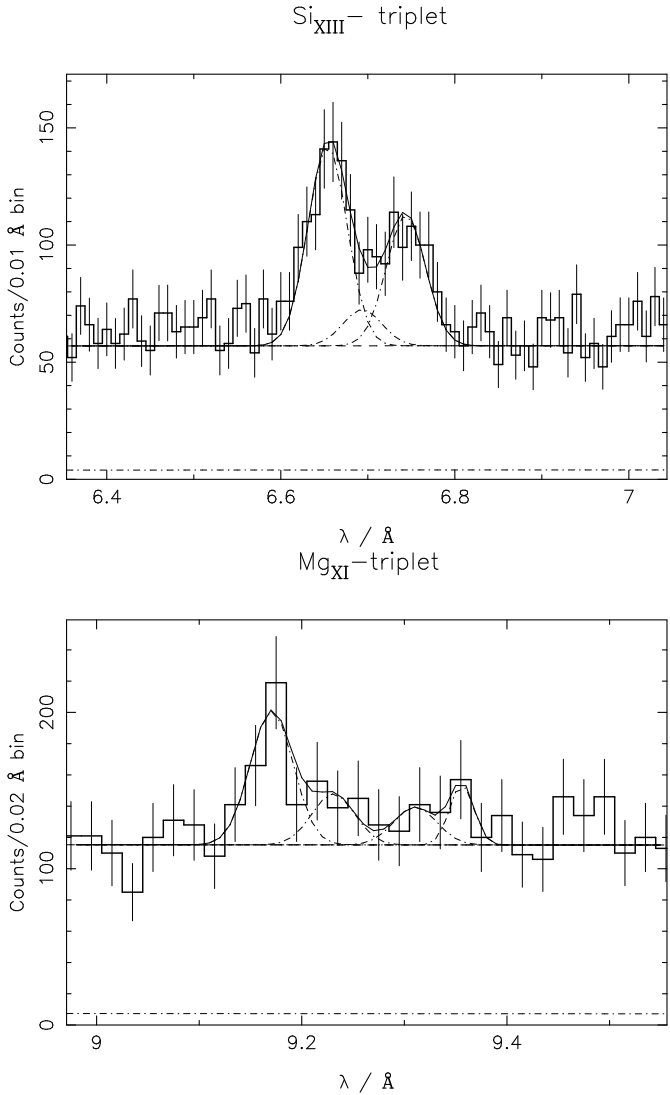

 \resizebox{\hsize}{!}{\rotatebox{270}{\includegraphics{MS2207f10}}}
 \resizebox{\hsize}{!}{\rotatebox{270}{\includegraphics{MS2207f11}}}
 \caption{ (a,b) Same as Fig.~\ref{spectra_on} for Si\,{\sc xiii} (a) and
Mg\,{\sc xi} (b). The unidentified line at 9.36\,\AA\ is also fitted with
57~cts.}
 \label{spectra_simgne}
\end{figure}

For the analysis of emission lines we use a maximum likelihood method
which compares the sum of a model and the instrumental background
with the non-subtracted count spectrum. In this way Poisson statistics can be
explicitly taken into account. The model spectrum consists of one or more lines
with a variable or fixed spacing and a source background, which is assumed
to be constant over the region of interest (i.e., the spectral lines under
individual consideration). This assumption is well justified since both the
instrumental background is quite flat as well as the source background
(bremsstrahlung spectrum) once multiplied with the effective areas. The method
includes Poisson fluctuations both in the line and background counts. Our code
assumes the line
profile functions to be Gaussian, but other shapes can be easily implemented.
In this paper we used the CORA program\footnote{detailed description under\\
{\tt http://www.hs.uni-hamburg.de/DE/Ins/Per/Ness/Cora}}, version 1.2, for
the analysis. It has been developed and described by \cite{ness01a} (2001a),
and can be downloaded from {\tt http://ibiblio.org/pub/Linux/science/astronomy/}.

The analysis was performed on the basis of the count spectrum. The measured
line counts are given in Tab.~\ref{tab_res} and we list the best fits of the
wavelengths $\lambda$, the Gaussian line-widths, the number of line photons,
and the source background $sbg$ measured in counts/\AA\ which is assumed
constant within the individual parts of the spectrum under consideration.
In the last column we list the effective areas (as provided by \cite{dpease00}
Oct. 2000) as used for calculating line ratios needed for further analysis from
the measurements; all errors in Tab.~\ref{tab_res} are 1$\sigma$ errors.
The first part of Tab.~\ref{tab_res} contains the He-like triplets
Si\,{\sc xiii}, Mg\,{\sc xi}, Ne\,{\sc ix}, O\,{\sc vii}, and N\,{\sc vi} in
combination with their H-like lines Si\,{\sc xiv}, Mg\,{\sc xii}, Ne\,{\sc x},
O\,{\sc viii}, and N\,{\sc vii}. Since the Ne triplet is severely blended,
the contaminating lines are also listed in Tab.~\ref{tab_res}.\\
For the density diagnostics of the higher temperature regions, five
Fe\,{\sc xxi} lines are also listed in Tab.~\ref{tab_res}, together with
the ratios of each line with respect to the Fe\,{\sc xxi} 128.73\,\AA\ line.
For an estimate of optical depth effects, two Fe\,{\sc xvii} lines were also
measured and further analysis is performed in Sect.~\ref{optdepth}.\\

\subsubsection{Analysis of He-like line ratios}
\label{heanal}

We first discuss the lower temperature He-like line systems from oxygen and
nitrogen. In Fig.~\ref{spectra_on} (a,b) we show the region around the O\,{\sc
vii} triplet at 22\,\AA\ and the N\,{\sc vi} triplet at 28\,\AA\ together with
best fits of the resonance, intercombination, and forbidden lines. All three
lines are clearly detected above the background, which is actually dominated by
continuum radiation from Algol itself (cf., Fig.~\ref{sp_all}).\\

In Fig.~\ref{spectra_simgne} (a,b) we show the Mg\,{\sc xi} and Si\,{\sc xiii}
triplets
together with our best fits. Obviously, the relative spectral resolution
of the LETGS becomes smaller with smaller wavelengths, and at short wavelengths
the {\it Chandra} HETGS performs far better. Still, the lines are at 
least partially resolved and line parameters can be determined by fitting
a line template to the data whose relative position is fixed. In this 
fashion the Si\,{\sc xiii} triplet blend can be fitted and the determined 
value for the f/i ratio is consistent with the low density limit. The
Mg\,{\sc xi} triplet is more complicated. While the Mg\,{\sc xi} r-line is
clearly detected, there are no clear detections of the i and f-lines. In
particular, the emission line feature(s) found at the expected position of the
Mg\,{\sc xi} f-line is unusually broad, yet we are not aware of
other strong contaminating lines in that region as is suggested by the fit in
Fig.~\ref{spectra_simgne} (b). The determined line fluxes and hence line
ratios do of course depend on the adopted background levels, yet in no case
do we find an f/i-ratio consistent with the low density limit. Since this
is in conflict with both the Si\,{\sc xiii} data as well as the Fe\,{\sc xxi}
data discussed below, we consider the ``detections'' of the Mg\,{\sc xi} i and
f lines shown in Fig.~\ref{spectra_simgne} (b) and reported in
Tab.~\ref{tab_res} as spurious.

\subsubsection{Analysis of the Ne\,{\sc ix} triplet}
\label{ne_anal}

\begin{figure}[!ht]
 \resizebox{\hsize}{!}{\rotatebox{270}{\includegraphics{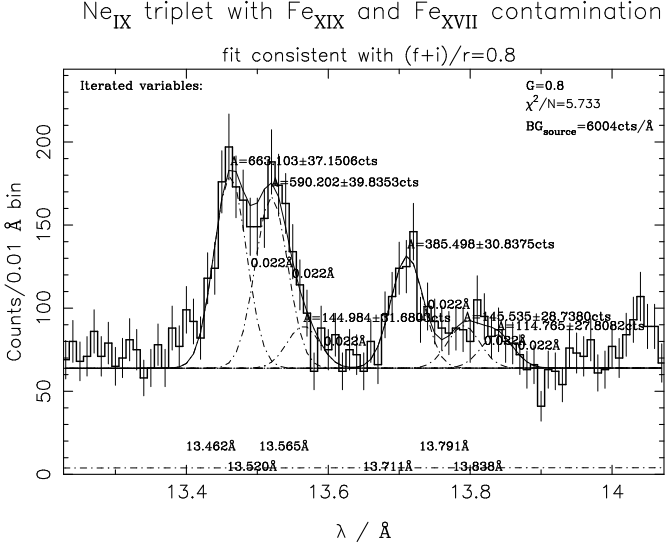}}}
 \caption[]{The Ne\,{\sc ix} triplet with blending lines from
Fe\,{\sc xix/xvii}. The fit is constrained to (f+i)/r=0.8, i.e., granting a
meaningfull G ratio, assuming log$T<6.7$ (5\,MK) for the Ne triplet.}
 \label{ne_500}
\end{figure}

The analysis of the Ne\,{\sc ix} triplet is notoriously difficult, because of
severe blending of the intercombination line at 13.55\,\AA\ with an
Fe\,{\sc xix}
line at 13.52\,\AA\ (cf., Fig.~\ref{ne_500} and Tab.~\ref{tab_res}). The
Ne\,{\sc ix} resonance and the forbidden lines are clearly detected, while the
intercombination line is ``lost'' in a large line blend longward of the
resonance line. Our fits indicate 665 counts in the r-line,
385 counts in the f-line, and 736 counts in the
i-line blend. The question is how many of those 736 counts are
due to the Ne\,{\sc ix} i-line rather than Fe\,{\sc xix}. In the following we
estimate that by, first, constraining the fit to G=(f+i)/r=0.8, and,
second, by extrapolating line fluxes from other Fe\,{\sc xix} lines.

The fit shown in Fig.~\ref{ne_500} was performed enforcing the boundary
condition of G=0.8. With this constraint a reasonable fit is obtained
with 147 counts in the intercombination line and the remaining 589 counts in the
Fe\,{\sc xix} line. We now discuss whether this line count of 589 counts is
consistent with extrapolations from other Fe\,{\sc xix} lines using MEKAL
(\cite{mewe95} 1995) for calculating flux ratios.
For this purpose we selected four other Fe\,{\sc xix} lines suitable
for comparison, i.e., they are sufficiently isolated and/or sufficiently
strong. These lines are located at 13.79\,\AA, 14.67\,\AA, 101.5\,\AA, and
108.5\,\AA, and our fit results are listed in Tab.~\ref{fe19_extra}; for
comparison we use as reference line the Fe\,{\sc xix} at 13.52\,\AA.
The measured flux ratios can be used for comparison with theoretical line flux
ratios taken from MEKAL (\cite{mewe95} 1995). From the temperature analysis in
Sect.~\ref{temp_diag} and from Tab.~\ref{ftemp} we assume a temperature of
10\,MK for extrapolating flux ratios from the theoretical fluxes.

\begin{table}
\caption[ ]{\label{fe19_extra}Measured line flux ratios for Fe\,{\sc xix} and
Fe\,{\sc xvii} lines used for consistency checks of the treatment of the
Ne\,{\sc ix} blend in Fig.~\ref{ne_500}.
Values for effective areas $A_{\rm eff}$ are taken from In-Flight Calibration by
\cite{dpease00} (31. October 2000). The measured ratios are corrected for
interstellar absorption with the values from Tab.~\ref{tab_res}.
Theoretical flux ratios are obtained from MEKAL (\cite{mewe95} 1995) assuming
log$T=7.0$ (10\,MK, cf. Tab.~\ref{ftemp}).}
\begin{flushleft}
\renewcommand{\arraystretch}{1.2}
\begin{tabular}{l|c|c||l|p{1.cm}|c}
\hline
$\lambda$/\AA&A [cts]&A/$\lambda$/$A_{\rm eff}$&&\multicolumn{2}{c}{Flux ratio}\\
&&$\sim$[erg/cm$^2$]&{\scriptsize $\lambda$/13.52}&meas.&theor.\\
\multicolumn{2}{l}{Fe\,{\sc xix} lines}& &13.79 & 0.24 $\pm$ 0.05 &0.30\\
13.52 & 588.98 & 1.66 $\pm$ 0.11 & 14.67 & $<$ 0.12 &0.11\\
13.79 & 147.30 & 0.40 $\pm$ 0.07 & 101.5 & 0.07 $\pm$ 0.02 &0.16\\
14.67 & $<$ 80. & $<$ 0.20 & 108.5 & 0.19 $\pm$ 0.02&0.42\\
\cline{4-6}
101.5 & 77.670 & 0.11 $\pm$ 0.02 & {\scriptsize $\lambda$/13.79}&&\\
108.5 & 204.00 & 0.29 $\pm$ 0.02 & 13.52 & 4.09 $\pm$ 0.85 & 3.3\\
\cline{1-3}
&&&14.67 & $<$ 0.49 &0.39\\
&&&101.5 & 0.3 $\pm$ 0.08 &0.56\\
\multicolumn{2}{l}{Fe\,{\sc xvii} lines}& & 108.5 & 0.80 $\pm$ 0.18 & 2.5\\
\cline{4-6}
15.00 & 1042.7 & 2.56 $\pm$ 0.09 & \multicolumn{3}{l}{Ratio for Fe\,{\sc xvii} $\lambda$/13.84}\\
13.84 & 114.20 & 0.31 $\pm$ 0.07 & 15.00& 8.16 $\pm$ 2.0 & 12.9\\
\hline
\end{tabular}
\renewcommand{\arraystretch}{1}
\end{flushleft}
\end{table}

The measured Fe\,{\sc xix} and Fe\,{\sc xvii} ratios listed in
Tab.~\ref{fe19_extra} indicate that the measured flux ratios are systematically
smaller than the theoretical ratios from MEKAL. This effect is more significant
when using the Chianti data base. But the general trend is quite convincing
indicating the Fe\,{\sc xix} lines at 13.52\,\AA\ and 13.79\,\AA\ to be well
modelled in Fig.~\ref{ne_500} and in Tab.~\ref{tab_res}. Also the results
for Fe\,{\sc xvii} at 13.84\,\AA\ seem to be realistic.

Eventually a reduction of Fe\,{\sc xix} at 13.52\,\AA\ would cure the difference
between measured and theoretical ratios, leading to a higher value of the
intercombination line. This would mean a higher G ratio (i.e., a lower
temperature) and a lower f/i ratio (i.e., a higher density). But this reduction,
in combination with a reduced Fe\,{\sc xix} at 13.79\,\AA\ flux,
would also require an enhancement of the Fe\,{\sc xvii} line at 13.84\,\AA\ in
order to retain the sum of the blend at 13.79\AA/13.84\,\AA, and the ratio of
this line with the very strong 15\,\AA\ line would become smaller, which is the
opposite effect of what would be intended with the reduction. We therefore trust
our results obtained from the constrained fit only fixing the G ratio to 0.8.

\subsubsection{Analysis of Fe\,{\sc xxi} line ratios}

\begin{figure}
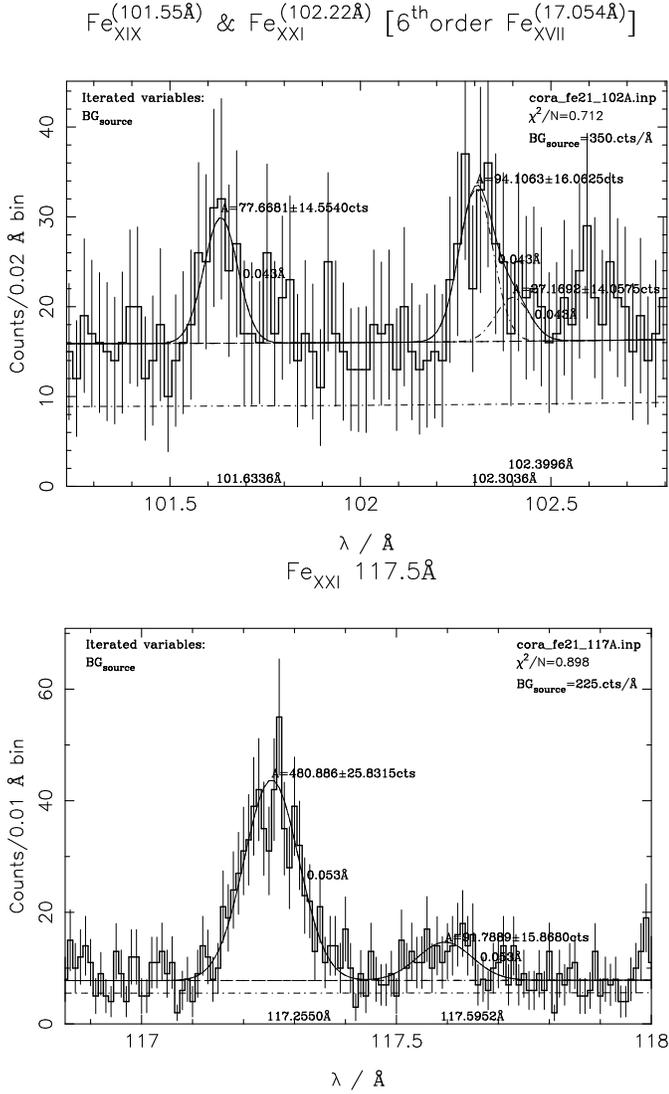

 \resizebox{\hsize}{!}{\rotatebox{270}{\includegraphics{MS2207f13}}}
 \resizebox{\hsize}{!}{\rotatebox{270}{\includegraphics{MS2207f14}}}
 \caption{Fitting of Fe\,{\sc xxi}
Top: Fe\,{\sc xxi} (102.22\,\AA) in combination with Fe\,{\sc xix}
(101.55\,\AA) and $6^{th}$ order of Fe\,{\sc xvii} (17.054\,\AA)
Bottom: Fe\,{\sc xxii} (117.17\,\AA) and Fe\,{\sc xxi} (117.505\,\AA)}
 \label{fe102}
\end{figure}

The analysis of most of the Fe\,{\sc xxi} lines was straightforward.
For the Fe\,{\sc xxi} line at 121.21\,\AA\ only an upper limit could be
determined. Some difficulties were encountered for the Fe\,{\sc xxi} 
102.22\,\AA\ and the
Fe\,{\sc xxi} 117.505\,\AA\ lines. The Fe\,{\sc xxi} line at 102.22\,\AA\
is partially blended with the 6$^{th}$ order of Fe\,{\sc xvii}
at the original wavelength at 17.054\,\AA. The Fe\,{\sc xxi} line at
117.505\,\AA\ is found to be very broad such that it is difficult to find the
correct wavelength position. In the top panel of
Fig.~\ref{fe102} our model with the isolated Fe\,{\sc xix} line at 101.55\,\AA\
together with Fe\,{\sc xxi} at 102.22\,\AA\ and the 6$^{th}$ order of
Fe\,{\sc xvii} is shown. In the bottom panel of Fig.~\ref{fe102} the
Fe\,{\sc xxi} at 117.505\,\AA\ is shown in combination with the strong, isolated
Fe\,{\sc xxii} line at 117.17\,\AA. In both cases the isolated lines are used to
determine the line shift of our measurement in comparison with the theoretical
wavelengths. In that way the expected wavelength position for the weaker, or
blended lines under consideration was used for the fit.

\begin{table*}[!ht]
\caption[ ]{\label{tab_res}Measured line counts for Algol with 1$\sigma$ errors.
Values of the total effective areas $A_{\rm eff}$ are taken from In-Flight
Calibration by \cite{dpease00} (31. October 2000). The Transmissions from the
ISM are based on N(H\,{\sc i}) = 2.5$ \times 10^{18}$, N(He\,{\sc i})/N(H\,{\sc
i}) = 0.09, N(He\,{\sc ii})/N(H\,{\sc i}) = 0.01}
\begin{flushleft}
\renewcommand{\arraystretch}{1.2}
\begin{tabular}{r|c|c|c|c|c|c|c}
&$\lambda$\ [\AA]&$\sigma$\ [\AA]&A\ [cts]&sbg&$R_{\rm obs}=$f/i&$G_{\rm obs}=\frac{\rm i+f}{\rm r}$& $A_{\rm eff}$\\
&&&& [cts/\AA]&$[R]^{(2)}$&$[G]^{(2)}$&[cm$^2$]\\
\hline
\hline
\multicolumn{5}{l|}{\it He-like and H-like (cf. Sects.~\ref{temp_diag} and \ref{he_diag})}&&\\
\hline
Si\,{\sc xiv}&6.19 $\pm$ 0.0016&0.025 $\pm$ 0.002&658.32 $\pm$ 35.06&5119&&&36.22\\
Si\,{\sc xiii}$r$&6.65 $\pm$ 0.004   &&480.7 $\pm$ 35.95 &&&&37.54\\
$i$&6.69 $\pm$ 0.014   &0.022$\pm$ 0.003 &86.4 $\pm$ 33.10&  5300 &3.64 $\pm$ 1.44&0.83 $\pm$ 0.15&37.38\\
$f$&6.74 $\pm$ 0.006   &&314.1 $\pm$ 30.60 &&[3.66 $\pm$ 1.45]$^{(2)}$&[0.84 $\pm$ 0.15]$^{(2)}$&37.16\\
\hline
Mg\,{\sc xii}&8.42 $\pm$  0.0016&0.02 $\pm$  0.002&578.04 $\pm$ 33.13&5900&&&32.28\\
Mg\,{\sc xi}$r$& 9.17 $\pm$ 0.0015  &   &  224.12 $\pm$ 26.15&&&&27.69\\
$i$& 9.23 $\pm$ 0.0015  &  0.02 $\pm$ 0.002  &   84.85 $\pm$ 23.29&5400&0.90 $\pm$ 0.36&0.72 $\pm$ 0.22&27.42\\
$f$& 9.31 $\pm$ 0.0015  &    &   76.09 $\pm$ 22.80&&[0.90 $\pm$ 0.37]$^{(2)}$&[0.73 $\pm$ 0.23]$^{(2)}$&27.24\\
\hline
Ne\,{\sc x}&12.14 $\pm$ 0.0004&0.021 $\pm$ 0.0008&2481.48 $\pm$ 56.31&6510&&&24.96\\
Ne\,{\sc ix}&{\it cf. Sect.\ref{ne_anal}}&&&&&&\\
$r$&13.46 $\pm$ 0.003  &  0.022 $\pm$ 0.0022 &   665.03  $\pm$ 37.21&&&&26.16\\
$i$& 13.56 $\pm$ 0.011  &  0.022 $\pm$ 0.0022 &  146.65  $\pm$ 31.74 &6000&2.63 $\pm$ 0.61 & 0.80&26.23\\
$f$& 13.71 $\pm$  0.004 &  0.022 $\pm$ 0.0022 &   385.38 $\pm$ 30.83 &&[2.62 $\pm$ 0.61]$^{(2)}$&--&26.29\\
Fe\,{\sc xix}& 13.52 $\pm$  0.003 &  0.022 $\pm$ 0.0022 &  588.98 $\pm$ 39.82 &&
\multicolumn{2}{l|}{[fit constrained to G=0.8]}&26.21\\
Fe\,{\sc xix} & 13.79 $\pm$ 0.007  & 0.022 $\pm$ 0.0022 &  147.29 $\pm$ 28.81 &&&&26.31\\
Fe\,{\sc xvii} & 13.84 $\pm$ 0.07  & 0.022 $\pm$ 0.0022 &114.18 $\pm$ 27.80 &&&&26.32\\
\hline
O\,{\sc viii}&18.9701 $\pm$ 0.0004&0.0262 $\pm$ 0.0004&2882.96 $\pm$ 57.81&3150&&&24.29\\
O\,{\sc vii}$r$&21.62 $\pm$ 0.016&0.022 $\pm$ 0.002&262.49 $\pm$ 22.6&&&&15.58\\
$i$&21.82 $\pm$ 0.021&0.021 $\pm$ 0.004&128.77 $\pm$ 18.7&2439&0.94 $\pm$ 0.2&0.95 $\pm$ 0.16&15.34\\
$f$&22.11 $\pm$ 0.022&0.020 $\pm$ 0.004&120.9 $\pm$ 18.0&&[0.94 $\pm$ 0.2]$^{(2)}$&[0.97 $\pm$ 0.16]$^{(2)}$&15.32\\
\hline
N\,{\sc vii}&24.8 $\pm$ 0.0012&0.03 $\pm$ 0.0012&1119.05 $\pm$ 38.38&&&&15.23\\
N\,{\sc vi}$r$&28.81 $\pm$ 0.007&0.037 $\pm$ 0.008&141.33 $\pm$ 21.10&&&&13.57\\
$i$&29.11 $\pm$ 0.011&0.058 $\pm$ 0.012&188.23 $\pm$ 25.53&1700&0.20 $\pm$ 0.08&1.60 $\pm$ 0.37&13.57\\
$f$&29.55 $\pm$ 0.011&0.021 $\pm$ 0.007&37.5 $\pm$ 14.31&&[0.21 $\pm$ 0.09]$^{(2)}$&[1.65 $\pm$ 0.38]$^{(2)}$&12.76\\
\hline
\hline
\multicolumn{5}{l|}{Fe\,{\sc xxi} \it density diagnostics (cf. Sect.~\ref{fe21_dens})}&
$\frac{\rm F(\lambda)}{\rm F(128.73\textrm{\tiny \AA})}^{(1)}$&ISM&\\
\hline
&97.87    & 0.049  &  67.20 $\pm$ 14.50  & 334 &0.12 $\pm$ 0.03&0.9206&7.17\\
&102.22   & 0.043  &  94.11 $\pm$ 16.1 & 350 &0.18 $\pm$ 0.07&0.9111&6.64\\
&117.505  & 0.058  &  92.10 $\pm$ 16.00  & 225 &0.20 $\pm$ 0.04&0.8734&6.13\\
&121.22   & 0.055  &  $<36$ & 117 &$<0.09$&0.8633&5.55\\
&128.73   & 0.058  &  266.32 $\pm$ 20.90 &  73 &1&0.8420&3.64\\
\hline
\multicolumn{5}{l|}{Ne\,{\sc ix} \it consistency check (cf. Sect.~\ref{ne_anal})}&&&\\
\hline
Fe\,{\sc xix} & 14.66 $\pm$ 0.007 & 0.02 $\pm$ 0.002 & $<$ 80 & 5010 &&0.999&26.99\\
Fe\,{\sc xix} & 101.63 $\pm$ 0.007 & 0.043 $\pm$ 0.006 & 77.67 $\pm$ 14.55 & 350 &&0.9124&6.69\\ 
Fe\,{\sc xix} & 108.45 $\pm$ 0.007 & 0.054 $\pm$ 0.006 & 203.53 $\pm$ 19.42 & 311 &&0.8964&6.42\\
\hline
\hline
\multicolumn{5}{l|}{Fe\,{\sc xvii} \it optical thickness (cf. Sect.~\ref{optdepth})}& $\frac{\rm F(15.03\textrm{\tiny \AA})}{\rm F(15.27\textrm{\tiny \AA})}$&
$\frac{\rm F(15.03\textrm{\tiny \AA})}{\rm F(15.27\textrm{\tiny \AA})}^{(2)}$&\\
\hline
Fe\,{\sc xvii}&15.026 $\pm$ 0.001 &  0.021 $\pm$ 0.001 & 1018.44 $\pm$ 38.93 &4500
&&&27.21\\
Fe\,{\sc xvii}&15.27 $\pm$ 0.008 &  0.023 $\pm$ 0.003 & 364.71 $\pm$ 28.92&4500&2.79 $\pm$ 0.25& [2.81 $\pm$ 0.25]$^{(2)}$& 27.42\\
\hline
\hline
\multicolumn{5}{l|}{Fe \it temperatures (cf. Sect.~\ref{temp_diag})}&${\rm A[cts/cm^2]}^{(2)}$&\\
\hline
Fe\,{\sc xvii} & 15.28 $\pm$ 0.004 & 0.024 $\pm$ 0.003 & 375.73 $\pm$ 29.30 & 4500& 14.22 $\pm$ 1.11 &
&26.42\\
Fe\,{\sc xviii} & 16.082 $\pm$ 0.0031 & 0.03 $\pm$ 0.004 & 572.78 $\pm$ 35.95 & 4500&  21.07 $\pm$ 1.32& & 27.19\\
Fe\,{\sc xxii} & 117.25 $\pm$ 0.093 & 0.053 $\pm$ 0.06 & 483.14 $\pm$ 25.91 & 225 & 78.56 $\pm$ 4.21 & & 6.15\\
Fe\,{\sc xx} & 118.81 $\pm$ 0.014 & 0.0623 $\pm$ 0.013 & 97.87 $\pm$ 16.98 & 225&16.26 $\pm$ 2.82 & & 6.02\\
Fe\,{\sc xx} & 121.98 $\pm$ 0.009 & 0.0675 $\pm$ 0.009 & 180.88 $\pm$ 19.23 & 138& 34.19 $\pm$ 3.64& &5.29\\
\hline
\end{tabular}
\\
$^{(1)}$ Fluxes accounting for $A_{\rm eff}$ and ISM as in last but one column.
$^{(2)}$ Fluxes corrected for $A_{\rm eff}$ listed in the last column.
\renewcommand{\arraystretch}{1}
\end{flushleft}
\end{table*}


\section{Temperature diagnostics}
\label{temp_diag}

\begin{table}
\caption{\label{ftemp} Temperature diagnostics from the flux ratios of
Ly$_{\alpha}$/He$_r$ for the ions Si\,{\sc xiv/xiii}, Mg\,{\sc xii/xi},
Ne\,{\sc x/ix}, O\,{\sc viii/vii}, and N\,{\sc vii/vi}. $T_{\rm M}$ denotes the
peak formation temperatures. Temperatures for various Fe ions are calculated
in the same manner by using theoretical line ratios from MEKAL.}
\begin{flushleft}
\renewcommand{\arraystretch}{1.8}
\begin{tabular}{l c c c}
\hline
	&Ly$_\alpha$/r           &$T$(H-He)      &$T_{\rm M}$\\
	&&[MK]&[MK]\\
\hline
\large $\frac{\rm Si\,\mathsc{xiv}}{\rm Si\,\mathsc{xiii}}$ & 1.54 $\pm$ 0.14 & 14.6 $\pm$ 0.5 &15.85/10.0\\
\large $\frac{\rm Mg\,\mathsc{xii}}{\rm Mg\,\mathsc{xi}}$ & 2.41 $\pm$ 0.31 & 10.4 $\pm$ 0.5&10.0/6.3\\
\large $\frac{\rm Ne\,\mathsc{x}}{\rm Ne\,\mathsc{ix}}$  & 4.33 $\pm$ 0.26 & 7.5 $\pm$ 0.2 & 5.62/3.98\\
\large $\frac{\rm O\,\mathsc{viii}}{\rm O\,\mathsc{vii}}$  & 8.03 $\pm$ 0.71 & 4.8 $\pm$ 0.2&3.16/2.2\\
\large $\frac{\rm N\,\mathsc{vii}}{\rm N\,\mathsc{vi}}$  & 8.37 $\pm$ 1.28 & 3.4 $\pm$ 0.2&2.0/1.4\\
\hline
\hline
   &&flux ratio  & $T$/MK  \\
\hline
\multicolumn{2}{l}{Fe\,{\sc xvii}/Fe\,{\sc xviii}}&  &\\
\multicolumn{2}{c}{15.265/16.078}& 0.67 $\pm$ 0.07 & 7.93 $\pm$ 0.43\\
\multicolumn{2}{l}{Fe\,{\sc xxi}/Fe\,{\sc xxii}}&  &\\
\multicolumn{2}{c}{117.51/117.17}& 0.19 $\pm$ 0.03 & 10.41 $\pm$ 0.71\\
\multicolumn{2}{l}{Fe\,{\sc xxi}/Fe\,{\sc xxii}}&  &\\
\multicolumn{2}{c}{121.83/117.51}&2.28 $\pm$ 0.46 & 10.38 $\pm$ 0.83\\
\multicolumn{2}{c}{118.66/117.51}&1.08 $\pm$ 0.27 & 10.73 $\pm$ 1.22\\
\hline
\end{tabular}
\renewcommand{\arraystretch}{1}
\end{flushleft}
\end{table}

We carry out temperature diagnostics using temperature sensitive
line ratios of Ly$_\alpha$/He$_r$ and of Fe\,{\sc y}/Fe\,${\mathsc y+1}$.
The results are listed in Tab.~\ref{ftemp}.
The ratios Ly$_\alpha$/He$_r$ were calculated from the line fluxes corrected
for effective areas, as listed in Tab.~\ref{tab_res}. We assume plasma
emissivities as calculated in the Codes MEKAL (\cite{mewe85} 1985;
\cite{mewe95} 1995) and SPEX (\cite{kaastra96} 1996) and compare the measured
ratios with the calculated emissivity ratios in order to derive line formation
temperatures. The results are listed in Tab.~\ref{ftemp} as $T$(H-He).\\

In addition we also investigate the temperature of Fe emitting layers with
various Fe flux ratios (cf. Tab.~\ref{ftemp} bottom). We used the photon fluxes
corrected for effective areas from Tab.~\ref{tab_res} and compared the ratios
with theoretical ratios derived with MEKAL and SPEX, in the same manner as
for the Ly$_\alpha$/He$_r$ ratios. The theoretical flux of the
Fe\,{\sc xviii} 16.078\,\AA\ was corrected (enhanced) by a factor of 2.14
following \cite{mewe01} (2001).\\
From this analysis we find a cooler component of 8\,MK, which is consistent with
the Ly$_\alpha$/He$_r$ result for Ne. We also find hotter plasma at 10.5\,MK in
which the highly ionized Fe ions are formed. The ratios and derived 
temperatures are listed in Tab.~\ref{ftemp}. Clearly, a multitude of
spectral components is present in the X-ray spectrum and we defer a
discussion of the admissible emission measure distributions to a 
forthcoming paper.\\


\section{Density diagnostics}

Estimates of coronal density can be obtained from the density 
sensitive f/i ratio of He-like triplets and from the Fe\,{\sc xxi} line 
ratios. The He-like N\,{\sc vi}, O\,{\sc vii}, and
Ne\,{\sc ix} ions probe the lower temperature components, while the
Mg\,{\sc xi}, Si\,{\sc xiii}, and the Fe\,{\sc xxi} ions are used to probe the
higher temperature components of the coronal plasma. The low-Z He-like 
ions are sensitive
at densities log($n_e$) between 9 and 12, while the high-Z He-like ions can
only be used for higher densities above log$(n_e)>$12. Lower densities
log$(n_e)>11$ at high
temperatures $\approx$~10\,MK can be diagnosed from the Fe\,{\sc xxi} ratios.

\subsection{He-like ions}
\label{he_theo}

\begin{figure}
 \resizebox{\hsize}{!}{\includegraphics{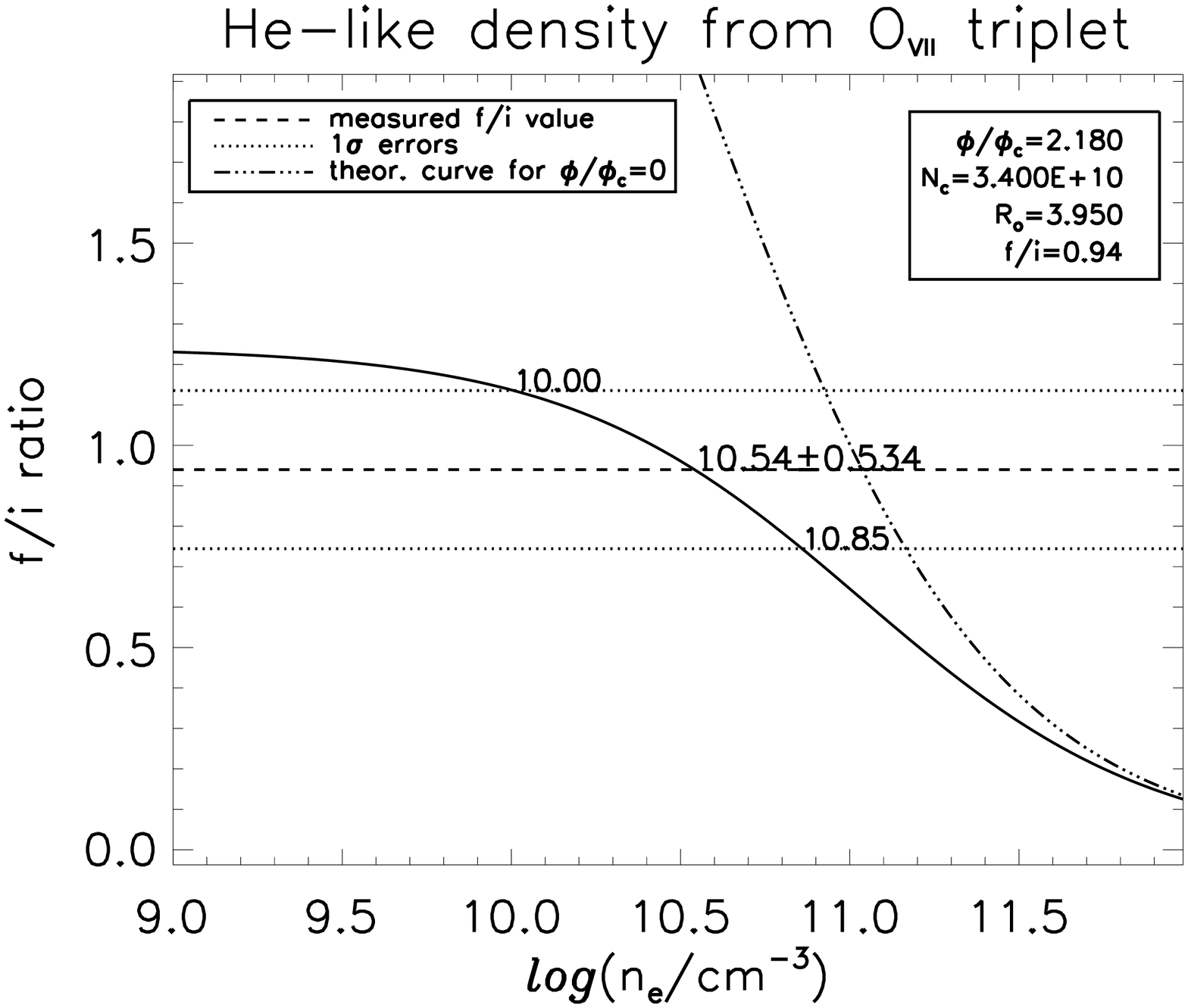}}
 \resizebox{\hsize}{!}{\includegraphics{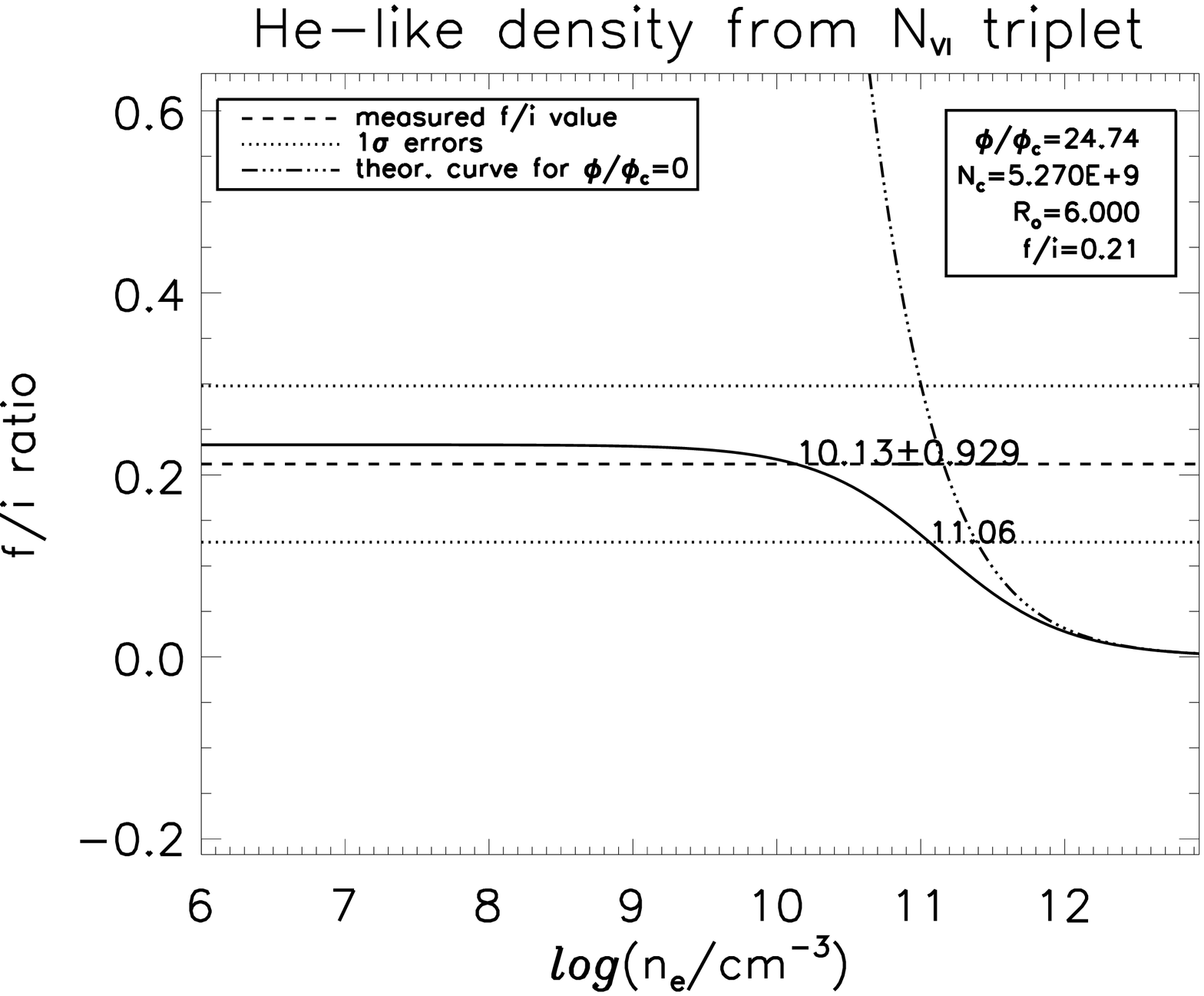}}
 \resizebox{\hsize}{!}{\includegraphics{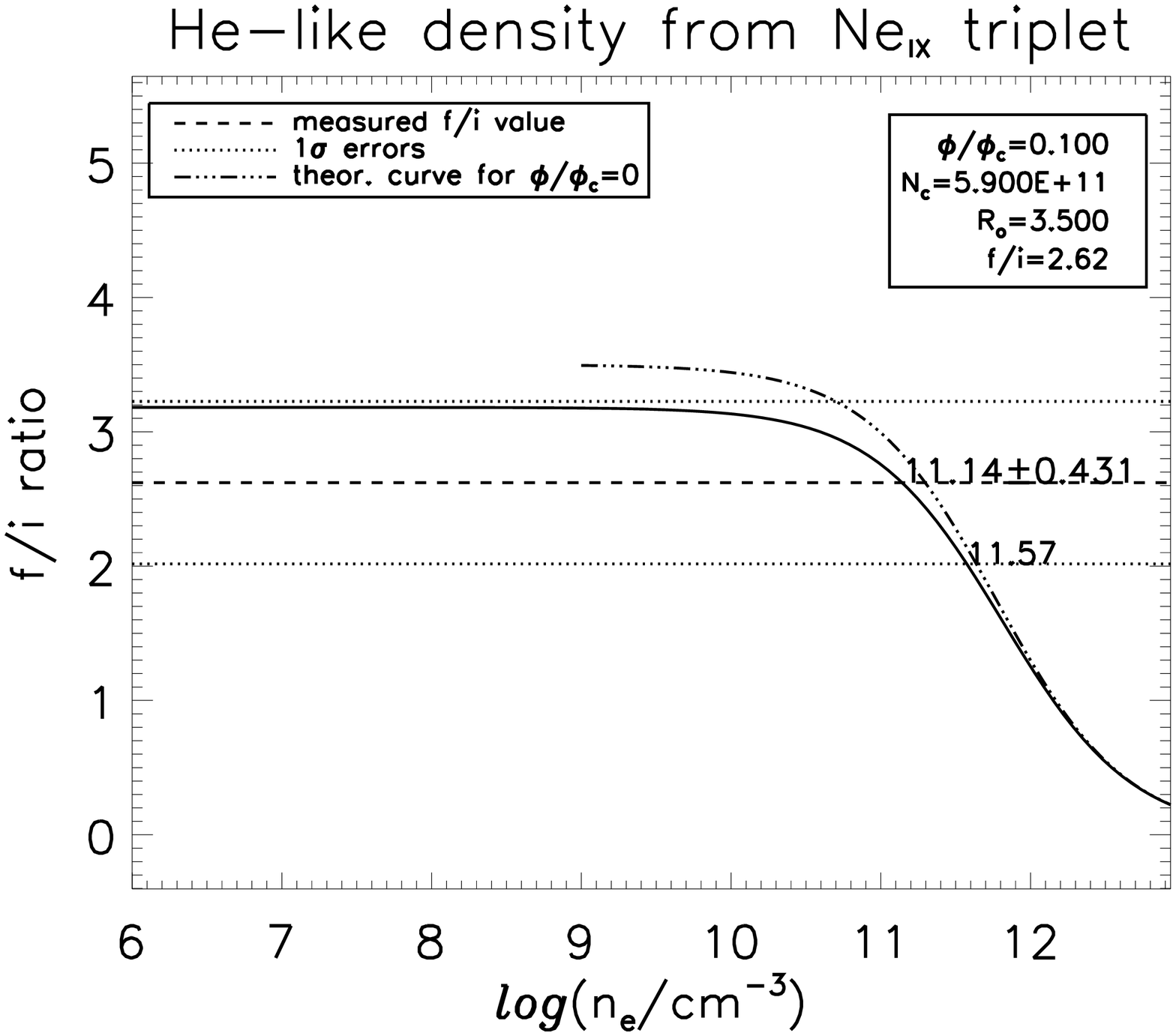}}
 \caption{\label{he_dens1}He-like densities $n_e$ from the ions O\,{\sc vii},
N\,{\sc vi}, and Ne\,{\sc ix}. For comparison the theoretical curve assuming
$\phi/\phi_c=0$ is also plotted (cf. Eq.~\ref{R_Ne}).}
\end{figure}

\begin{figure}
 \resizebox{\hsize}{!}{\includegraphics{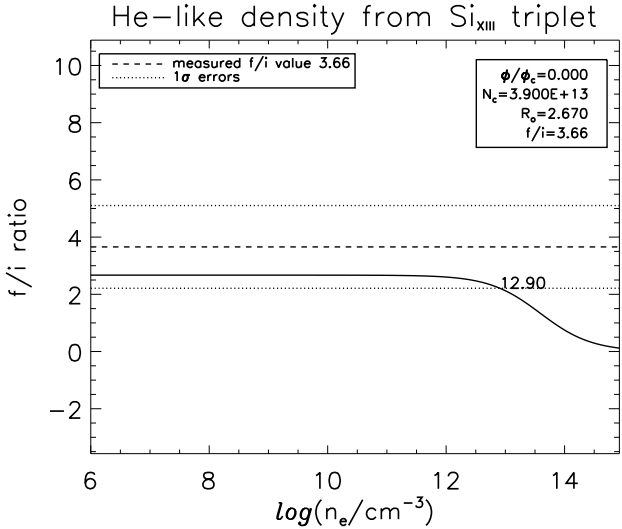}}
 \caption{\label{he_dens2}He-like densities $n_e$ from the Si\,{\sc xiii} ion.}
\end{figure}

\subsubsection{Theory of He-like triplets}

The theory of the atomic physics of He-like triplets has been extensively
described in the literature (\cite{gabriel69} 1969, \cite{blume72} 1972,
\cite{mewe78} 1978, \cite{pradhan81a} 1981, \cite{pradhan81b} 1981,
\cite{pradhan82} 1982, \cite{pradhan85} 1985, and recently
\cite{podub00} 2000, \cite{porquet00} 2001, and \cite{ness01a} 2001a).

In this paper we will determine electron densities $n_e$ from the equation
\begin{equation}
\label{R_Ne}
R(n_e) \ = \frac{\rm f}{\rm i}\ =
 \ \frac{R_0} {1 + \phi/\phi_c + n_e/N_{\rm c}}\,
\end{equation}
where $R_0$ denotes the low density limit and the parameter $N_{\rm c}$ the
so-called critical density (cf. Tab.~\ref{ionvals}), around which the observed
line ratio is density-sensitive.
Finally, the parameters $\phi$ (the radiative absorption rate from $2\,^3S$ to
$2\,^3P$ induced by an external radiation field) and $\phi_c$ describe the
additional possible influence of the stellar radiation field on the depopulation
of the $^3S$ state (cf., Sect.~\ref{rad_field}). Values for $R_0$ and
$N_{\rm c}$ used in this paper are listed in Tab.~\ref{ionvals}. Also listed
are the peak formation temperatures $T_{\rm M}$ for the ions.
\begin{table}
\caption[ ]{\label{ionvals}Atomic Parameters for He-like triplets.
$T_{\rm M}$ is the peak line formation temperature (MEKAL),
$R_0$ is the low-density limit and $N_{\rm c}$ is the density
were $R$ falls to half its low-density value. $R_0$ and $N_{\rm c}$
are taken from \cite{pradhan81b} (1981) and \cite{blume72} (1972).}
\begin{flushleft}
\renewcommand{\arraystretch}{1.2}
\begin{tabular}{r r r r}
\hline
ion & $T_{\rm M}$/MK & $R_0$ & $N_{\rm c}/(10^{10}$\,cm$^{-3})$\\
Si\,{\sc xiii}  &  10.0        &  2.67  &  3900 \\
Mg\,{\sc xi}  &  6.3        &    2.6  &  620 \\
Ne\,{\sc ix}  &  4.0     & 3.5         &   59.0\\
O\,{\sc vii} &   2.2       & 3.95       &   3.40\\
N\,{\sc vi}  &   1.4        & 6.0         &   0.53\\
\hline
\end{tabular}
\renewcommand{\arraystretch}{1}
\end{flushleft}
\end{table}

\subsubsection{Influence of the stellar radiation field}
\label{rad_field}

Given the effective temperature of 13000\,K of Algol~A and its close
proximity to Algol~B (cf., Tab.~\ref{star_prop}), we must check to what
extent the X-ray radiation originating from the
corona of Algol~B is influenced by the UV radiation from Algol~A;
the UV-radiation from the Algol~B component itself is small when comparing
its effective temperature with Capella ($T_{\rm eff}\leq$5000\,K) and 
the values computed for $\phi/\phi_c$ to be used in Eq.~\ref{R_Ne}
as derived by \cite{ness01a} (2001a). Also the measurements of other G and K
type stars, as presented by \cite{ness01c} (2001c) suggest Algol~B not to
contribute to the total radiation field. Following \cite{ness01a} (2001a) we
used IUE measurements of Algol in order to derive radiation temperatures for
the desired wavelengths, listed in Tab.~\ref{rad_temp}. From this we calculated
values $\phi/\phi_c$ for N\,{\sc vi}, O\,{\sc vii}, and Ne\,{\sc ix}
using a dilution factor of
\begin{equation}
W=\frac{1}{2}\left[1-\left\{1-\left(\frac{r_\star}{a}\right)^2\right\}^{1/2}\right]=0.01
\end{equation}
(\cite{mewe78} 1978) using as distance between Algol~A and B $a=14.6\,R_{\sun}$
(\cite{forb97} 1997) and the radius of Algol~A $r_\star=3.5\,R_{\sun}$
(Tab.~\ref{star_prop}).
The result of our analysis is listed in Tab.~\ref{rad_temp}. Since we assumed
the worst case scenario, we find the maximally possible values $\phi/\phi_c$,
so that the effects are not negligible for O\,{\sc vii} and not zero even
for Ne\,{\sc ix}. We do point out, however, that
in the phases between 0.74 and 1.06 (cf. Sect.~\ref{lc}) much of the visible
coronal emission could originate from regions not illuminated by the B star.
With an inclination angle of $81.5^\circ$ and the assumption of a uniform
distribution of the coronal plasma, geometrical considerations lead to the
result that at phase $\phi=1$ only plasma near the polar regions can be
illuminated by the primary component. A detailed discussion of these geometrical
considerations is given in \cite{ness01b} (2001b). In the following analysis we
calculate coronal densities with and without the effects of the radiation field
from the B star, i.e., assuming $\phi/\phi_c=0$ and with the values for
$\phi/\phi_c$ from Tab.~\ref{rad_temp}.\\

\begin{table}
\caption[ ]{\label{rad_temp}Investigation of the influence of the
stellar radiation field originating from the B8 star.
Measured fluxes from the IUE satellite $F_{\lambda}$
are converted to intensity $I_{\lambda}$ taking into account limb darkening
effects using $\epsilon=0.44$ for the U band (\cite{diaz} 1995).}
\begin{flushleft}
\renewcommand{\arraystretch}{1.2}
\begin{tabular}{r r r r}
\hline
&N\,{\sc vi}&O\,{\sc vii}&Ne\,{\sc ix}\\
\hline
$\lambda_{\rm f\rightarrow i}$/\AA&1900&1630&1266\\
$I_{\rm pot}$/eV&552.1&739.3&1195.3\\
$\frac{F_{\lambda}}{(10^{-10}\frac{\rm ergs}{\rm cm^{2}\,s\,
\textrm{\tiny \AA}})}$&$20\pm2$&$22\pm2.2$&$20\pm6$\\
$\frac{I_{\lambda}}{(10^7\frac{\rm ergs}{\rm cm^{2}\,s\,
\textrm{\tiny \AA}\,strd})}$&$9.1\pm0.91$&$9.98\pm1.0$&$9.1\pm2.7$\\
$T_{\rm rad}$/K&$12066\pm186$&$12708\pm190$&$13686\pm560$\\
dilution factor&\multicolumn{3}{c}{0.01}\\
$\phi/\phi_c$  & 24.74 $\pm$ 2.41& 2.18 $\pm$ 0.29&0.1 $\pm$ 0.03\\
\hline
\end{tabular}
\renewcommand{\arraystretch}{1}
\end{flushleft}
\end{table}

\subsubsection{Densities with the He-like ions}
\label{he_diag}

The measured ratios f/i, corrected for $A_{\rm eff}$ (\cite{dpease00} Oct. 2000;
the values are listed in the last column of Tab.~\ref{tab_res}), as quoted in
Tab.~\ref{tab_res}, were used for density diagnostics.
The theoretical curves from Eq.~\ref{R_Ne}, with the values $\phi/\phi_c$ from
Tab.~\ref{rad_temp} and the other parameters from Tab.~\ref{ionvals}, are
plotted in Figs.~\ref{he_dens1} and \ref{he_dens2} for each ion in
comparison with our measurements for f/i with 1$\sigma$ errors. For the low-Z
elements O, N, and Ne we also considered the case of no radiation field
from the B star
affecting the emitting layers with a line-dotted line in Fig.~\ref{he_dens1}.
The influence is most severe for N\,{\sc vi} and O\,{\sc vii}, but is also
visible for Ne\,{\sc ix}. Assuming a negligible radiation field from the
primary component we obtain definite deviations from the low density
limit for N\,{\sc vi} and O\,{\sc vii} and a marginal deviation for Ne\,{\sc ix}; in the latter
the latter case the 2$\sigma$ error includes the low density case. In
these cases we find densities between 1--2 $\times 10^{11}$\,cm$^{-3}$
(cf., Tab.~\ref{dens_tab}). Assuming the full radiation field supplied from the
B star to be effective the sensitivity of our measurements to detect densities
is significantly reduced because the difference between the R-value
in the high density case (R = 0) and low-density case becomes smaller and
smaller. Specifically, for our Algol LETGS spectrum we find that the data are
consistent with the low-density limit for nitrogen and neon, and even for
oxygen the low density limit is included within the 2$\sigma$ error bars.
For silicon and magnesium radiation effects are unimportant. The derived
f/i-ratio for silicon is consistent with the low density limit, while our
measurements for magnesium (formally) yield densities of $\approx$ log\,n$_e
= 13$; as discussed in Sect.~\ref{heanal} and Fig.~\ref{spectra_simgne} (b),
we consider the measurements of the i and f lines in magnesium spurious.


\subsection{Density diagnostic with Fe\,{\sc xxi}}
\label{fe21_dens}

In addition to He-like ions, ions with more than two electrons can be
used as a density diagnostic. The ground configuration of Fe\,{\sc xxi}
1s$^2$2s$^2$2p$^2$ splits up into $^3P$, $^1D$, $^1S$.
The energy difference between the ground state $^3P_0$
and the excited levels $^3P_1$ and $^3P_2$ is 9~eV and 14.5~eV,
respectively, and 30~eV between ground state and $^1D_2$.
In low-density plasmas virtually all atoms are in the ground state $^3P$, while
in high-density plasmas a Boltzmann equilibrium with the higher level will be
obtained. Consequently, from excited levels certain lines will appear only in
high-density plasmas. In contrast to He-like lines the appearance of certain
lines is an indicator of high-density plasmas.

\begin{figure}
 \resizebox{\hsize}{!}{\includegraphics{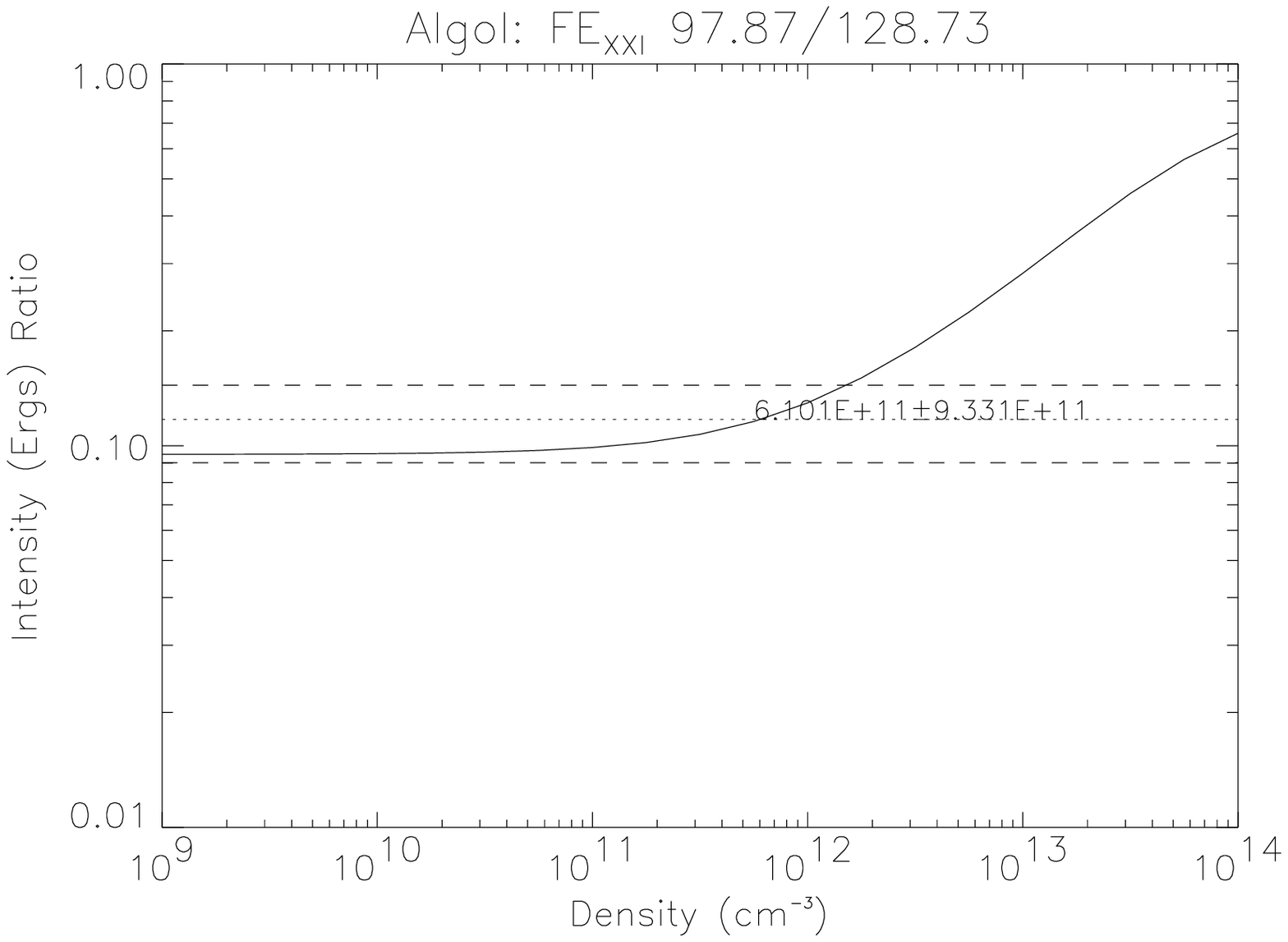}}
 \resizebox{\hsize}{!}{\includegraphics{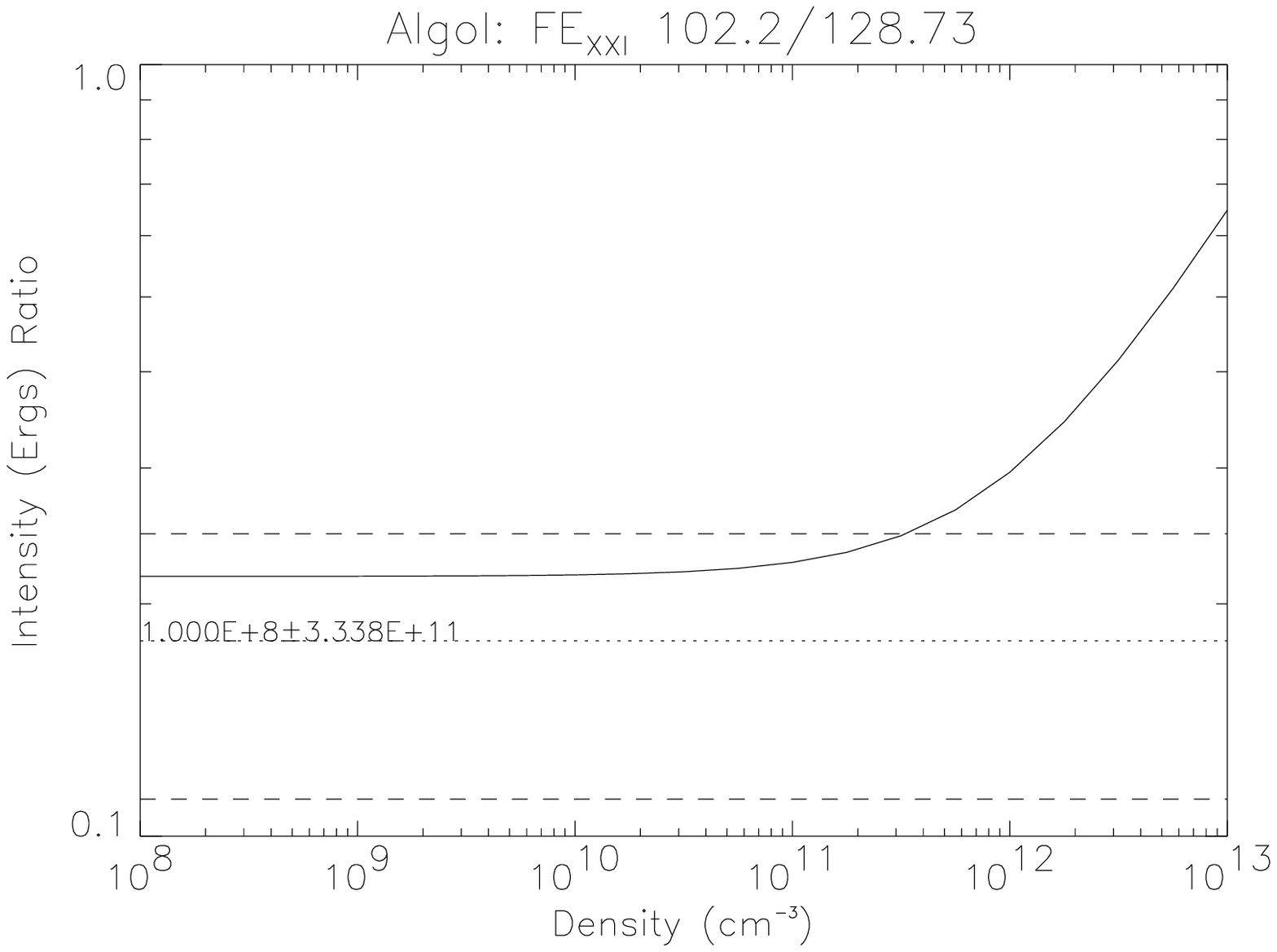}}
 \caption{Density diagnostic with Fe\,{\sc xxi} ions. 1$\sigma$ errors are
indicated with dashed line style (used 2$\sigma$ error for 102\,\AA~ratio).}
 \label{fe_dens}
\end{figure}

\begin{figure}
 \resizebox{\hsize}{!}{\includegraphics{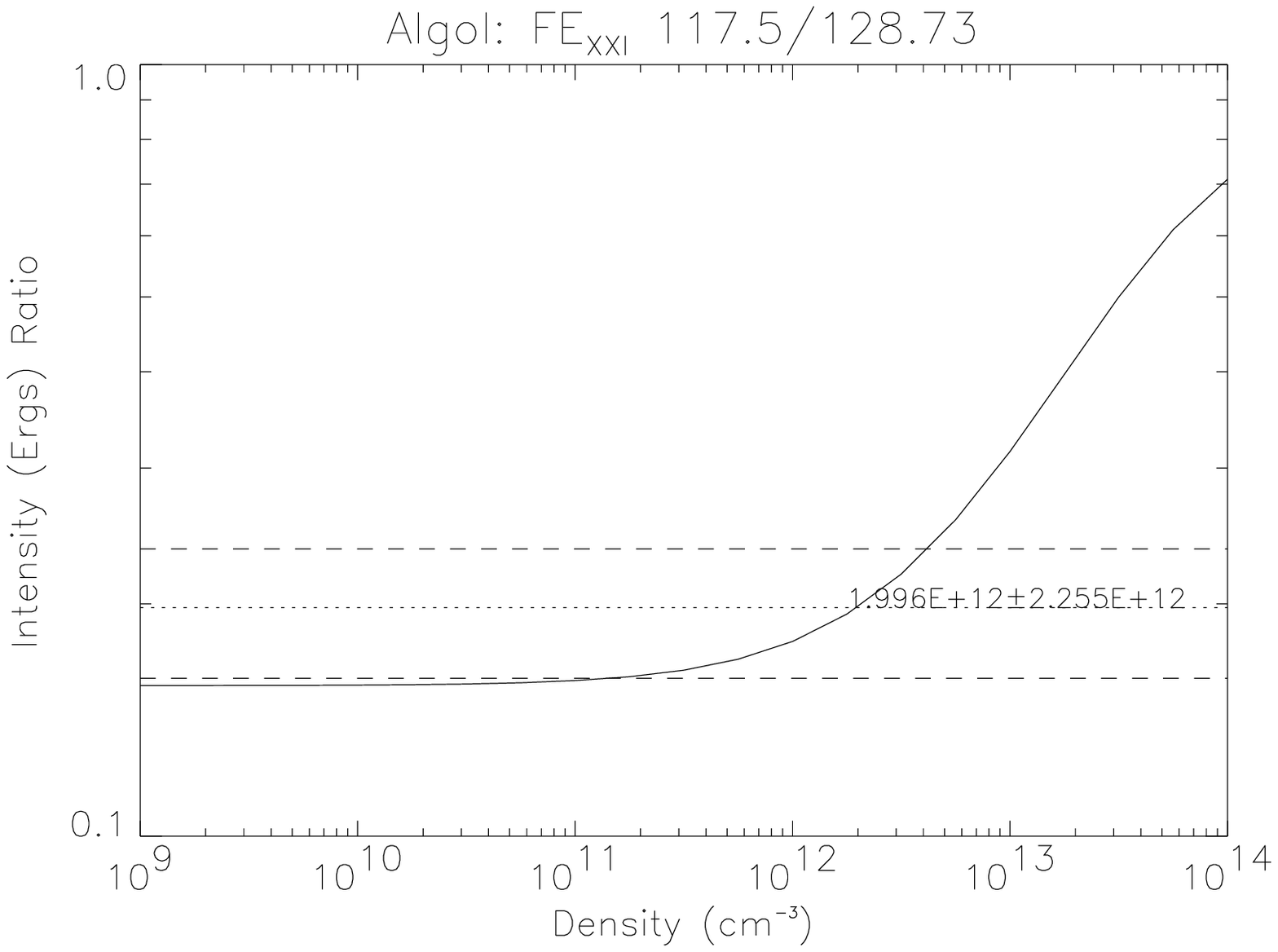}}
 \resizebox{\hsize}{!}{\includegraphics{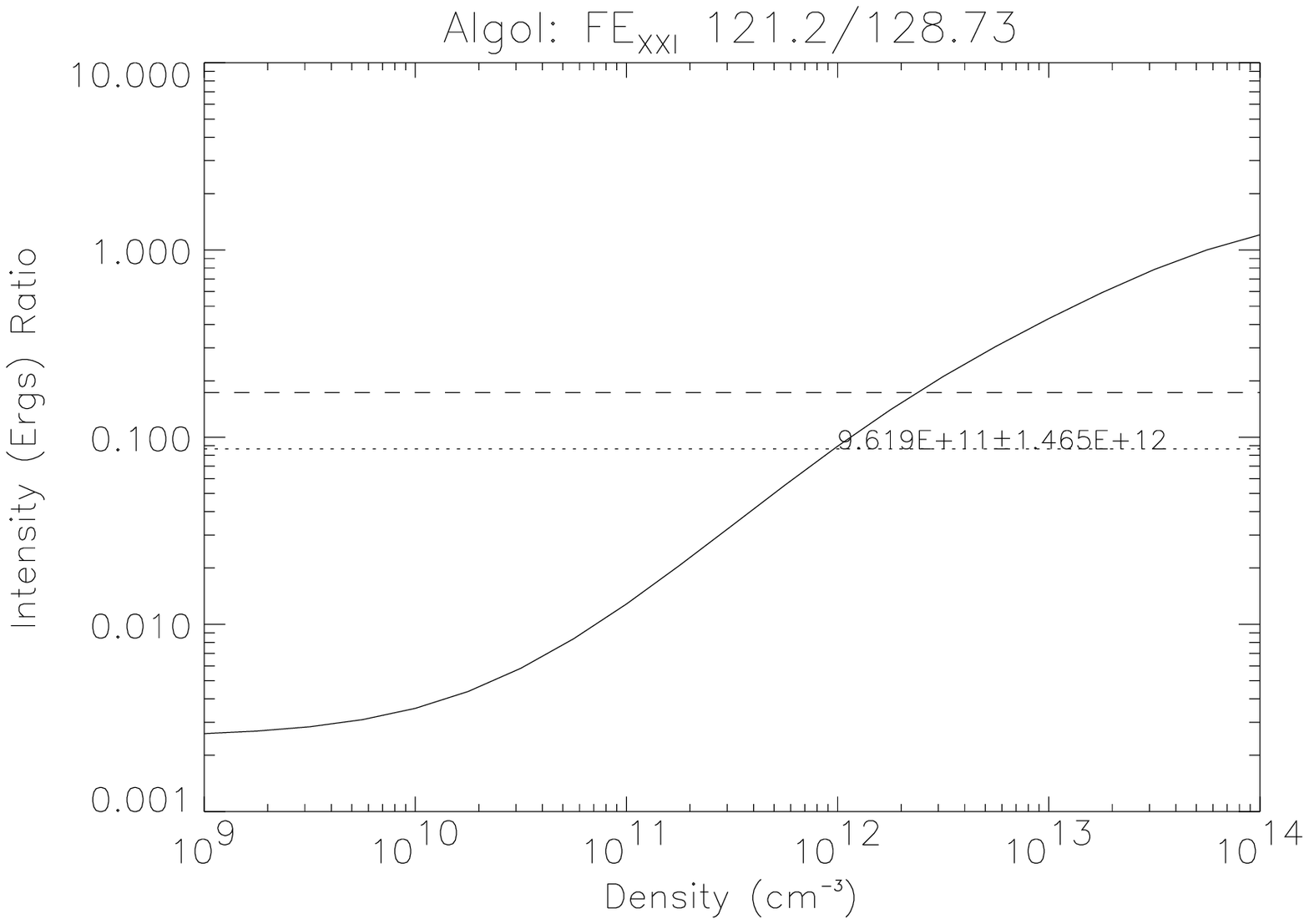}}
 \caption{Density diagnostic with Fe\,{\sc xxi} ions.}
 \label{fe_dens2}
\end{figure}

Our measurements of four Fe\,{\sc xxi} ratios, corrected for effective areas and
interstellar absorption are listed in Tab.~\ref{tab_res} (effective areas and
values from interstellar absorption are listed in the last two columns)
and plotted in comparison with theoretical flux ratio vs. $n_e$ curves obtained
with Chianti (\cite{dere01} 2001) in Figs.~\ref{fe_dens} and \ref{fe_dens2}. As
can be seen from Figs.~\ref{fe_dens} and \ref{fe_dens2}, only low density 
limits or upper limits are obtained for all Fe\,{\sc xxi} line ratios. Similar
results are obtained when using the ratios from \cite{brick95} (1995). The 
most sensitive upper limit (log\,n$_e < 11.52$) comes from
the Fe\,{\sc xxi} 102.2\,\AA/128.7\,\AA\ ratio, and is a factor $\sim$ 24 below
the upper limit derived for the Si\,{\sc xiii} triplet, which is formed at
similar temperatures. This comparison clearly shows that Fe\,{\sc xxi} line
ratios yield far more sensitive density constraints at high temperature as
compared to He-like triplets from magnesium, silicon, and higher ions.



\section{Optical depth effects}
\label{optdepth}

The {\it Chandra} Algol spectrum contains a number of strong
Fe\,{\sc xvii} lines which are sensitive to optical depth effects. 
In order to estimate the optical depth, we use an ''escape factor'' model with
a homogeneous mixture of emitters and absorbers in a slab geometry (e.g.,
\cite{kaame95} 1995, \cite{mewe01} 2001). In this geometry
the escape factor is $P(\tau) \sim [1+0.43\tau]^{-1}$. Significant
optical depths will lead to resonant scattering in strong lines.

If one considers two lines 
produced by the same ion, the ratio of the optical depths is given by
\begin{equation}
\label{tau1}
\frac {\tau _1} {\tau _2}  = \frac {\lambda _1 \ f_1}  {\lambda _2 \ f_2}.
\end{equation}
Therefore, one must look for lines from the same ion and large difference
in oscillator strength or wavelength. In the solar context one usually studies
the ratio of the strong Fe\,{\sc xvii} 15.03\,\AA\ resonance line ($f=2.66$) and
another, nearby Fe\,{\sc xvii} line with a small oscillator strength (e.g.,
15.265\,\AA, $f=0.593$). Estimating the escape factor $P$ from
$P=\left(\frac{15.03~{\mathrm \AA}}{15.27~{\mathrm \AA}}\right)_{\rm meas}/
\left(\frac{15.03~{\mathrm \AA}}{15.27~{\mathrm \AA}}\right)_{\tau=0}$
we can deduce the optical depth.

For the measured Fe\,{\sc xvii} 15.03/15.265 photon flux ratio we obtain a
formal fit result of 2.81 $\pm$ 0.25 (cf. Tab.~\ref{tab_res}). This ratio
does depend on the assumed background value and can vary between 2.32 and 3.45
when varying the source background (between 4000 and 5000 cts/\AA; comp.
Tab.~\ref{tab_res}). In this particular wavelength region (cf.,
Fig.~\ref{contspec}, middle) line blending is severe and a correct ``eye''
placement of the continuum is difficult.
However, our continuum modeling predicts relative stable values of
4500 cts/\AA\ so that we are confident that our quoted result is correct and
not affected by systematic errors from the placement of the continuum
background. We point out that this measurement agrees remarkably well with
the same line ratio as measured in Capella ({\cite{mewe01} 2001); this
is interesting because in Capella no measurements of density could be
obtained.

The measured photon flux ratio must be compared to the $\tau = 0$ flux ratio,
which can be deduced either from theory or laboratory measurements.
With SPEX we predict a ratio of 3.5, thus $P=(2.81\pm0.25)/3.5=0.8\pm 0.07$,
and therefore $\tau=0.57\pm 0.26$. With Chianti (\cite{dere01} 2001), a ratio of
$\approx 4$ is expected for $\tau=0$, and using $P=(2.81\pm0.25)/4=0.7\pm 0.06$
we find $\tau=0.98 \pm 0.3$. In either case we find optical depths
significantly different from zero (at $\sim 2-3 \sigma$ level).

Unfortunately theory does not agree with experiment. The very same line ratio
can be measured in the Livermore Electron Beam Ion Trap (EBIT; \cite{brown01}
2001, \cite{laming00} 2000). These experiments typically yield 
Fe\,{\sc xvii} 15.03/15.265 photon flux ratios in the range 2.5 - 3.0, which are
significantly different from those expected theoretically. Also, \cite{brown01}
(2001) point out that contamination of the 15.265\,\AA\ with Fe\,{\sc xvi}
further lowers the observed 15.03/15.265 photon flux ratio. Comparing the Algol
(and Capella) flux ratios in 15.03/15.265 to the values quoted by \cite{brown01}
(2001) we therefore conclude that the observations are fully consistent with an
optical thin plasma without any significant optical depth with possibly some
contamination arising from Fe\,{\sc xvi}. It is worrying that the theoretically
predicted emission from some of the strongest emission lines observed in solar
and stellar X-ray spectra appears to be wrong by $\approx$ 30 percent.

\begin{table}
\caption[ ]{\label{dens_tab}Summary of derived densities.\\
$^\star$ Resulting density values accounting for stellar radiation fields.\\
$^{\star\star}$ 2$\sigma$ upper limit.}
\begin{flushleft}
\renewcommand{\arraystretch}{1.2}
\begin{tabular}{l l l}
\hline
\multicolumn{3}{c}{He-like triplets}\\
\hline
&f/i&log($n_e$)\\
Si\,{\sc xiii}&3.66 $\pm$ 1.44&$<$12.9\\
Mg\,{\sc xi}&0.90 $\pm$ 0.37&n.a.\\
Ne\,{\sc ix}&2.62 $\pm$ 0.61&11.30 $\pm$ 0.60\\
$^\star$Ne\,{\sc ix}&&$<$11.57\\
O\,{\sc vii}&0.94 $\pm$ 0.20&11.04 $\pm$ 0.13\\
$^\star$O\,{\sc vii}&&10.54 $\pm$ 0.53\\
N\,{\sc vi}&0.21 $\pm$ 0.09&11.16 $\pm$ 0.23\\
$^\star$N\,{\sc vi}&&10.13 $\pm$ 0.93\\
\hline
\multicolumn{3}{c}{Fe\,{\sc xxi} ratios}\\
\hline
$\lambda/$\AA&ratio&log($n_e$)\\
97.87&0.12 $\pm$ 0.03&$<$12.19\\
102.22&0.18 $\pm$ 0.03&$<11.52^{\star\star}$\\
117.51&0.20 $\pm$ 0.04&$<$12.63\\
121.22&$<$0.09&$<$12.4\\
\hline
\end{tabular}
\renewcommand{\arraystretch}{1}
\end{flushleft}
\end{table}

\section{Discussion and Interpretation}
\label{loopmod}
From the derived parameters we can obtain structural information
on Algol's corona. From the lower temperature He-like ions
definite coronal densities could be determined, from the higher temperature
He-like ions as well as from the Fe\,{\sc xxi} line ratios (cf.,
Tab.~\ref{dens_tab}) upper limits to the coronal density at a temperature of
10\,MK can be derived. Since EM = $n_e^2\,V$, typical coronal volumes $V$
can be estimated from the recorded emission measures.\\

We now assume that Algol's corona is composed of a multitude of individual but
identical loops, all of which obey the loop scaling equation (\cite{rtv78} 1978)
\begin{equation}
\label{rtv}
n_e\,L=1.3\times 10^6\,T_{\rm apex}^{\, 2},
\end{equation}
with the densities $n_e$, the apex temperature $T_{\rm apex}$, and the loop
semilength $L$, all in $cgs$ units. Of course, the apex temperature $T_{\rm
apex}$ cannot be directly measured, it must however be close to the formation
temperature of the ion with the highest ionization temperature and it can
certainly not exceed the temperature of the bremsstrahlung continuum
(Sect.~\ref{cont}).
It is also clear that in Algol one is most likely dealing with a distribution
of X-ray emitting loops as is the case in the solar corona. However,
our data does not allow us to constrain the parameters of these
distributions and therefore we model the distributions by their presumed
mean values. From Tab.~\ref{ftemp} we conclude that the apex temperatures
$T_{\rm apex}$ must be approximately 15\,MK, which is also supported by the
continuum temperature, while from Tab.~\ref{dens_tab} we
conclude that log\,$n_e < $ 11.5. Obviously the densities at 15\,MK could be
arbitrarily low. However, if we assume that the material at 15\,MK is
in thermal pressure equilibrium with the material at 2\,MK (where the
O\,{\sc vii} line is produced) we estimate densities in the range
log\,$n_e$ $\sim$ 9.7 - 10.2, depending on whether the X-ray emitting corona
of Algol~B is immersed in the UV radiation field of Algol~A or not.
For log\,$n_e = $ 11.5 we compute $L = 9\times 10^8$\,cm while for densities
log\,$n_e$ = 9.7 - 10.2 we find $L = 180 - 460\times 10^8$\,cm. In either case,
the loop lengths are much smaller than the stellar radius of 2.45 $\times
10^{11}$\,cm so that we assume to deal with an essentially planar geometry.
If we assume ``canonical loops'' with circular cross sections of a tenth
of the loop length, the volumes of such coronal building blocks are
in the range 2.5\,10$^{25}$\,cm$^3$ to 3\,10$^{30}$\,cm$^3$, again
depending on the assumed density, and in all cases one requires at
least 1000 loops (or more) in order to account for the total observed
emission measure.

Let us next assume that the considered loops are semicircular and extend
to height
\begin{equation}
H = \frac {2 L} {\pi}. 
\end{equation}
We define the coronal filling factor as the ratio between the available volume
$4\pi R^2H$ and the actual volume of X-ray emitting material $V_{\rm cor}=
EM/n_e^2$:
\begin{equation}
f = \frac {V_{\rm cor}} {V_{\rm available}} = \frac {EM} {8 n_e^2 R^2 L},
\end{equation}
where $R$ denotes the stellar radius. Inserting the scaling law to replace $L$
with $T_{\rm apex}$ and $n_e$, we finally find
\begin{eqnarray}
\label{feq}
f= 9.6\times 10^{-8} \frac {EM} {n_e R^2 T_{\rm apex}^2}=
9.6 \frac{EM_{52}}{n_{10}R_{11}^2T_6^2}\\
\mbox{\footnotesize[$EM_{52}=EM/10^{52}\,{\rm cm}^{-3}, R_{11}=R/10^{11}\,{\rm cm},$}&&\nonumber\\
\mbox{\footnotesize{$T_6=T_{\rm apex}/10^6$\,K$, n_{10}=n_e/10^{10}\,{\rm cm}^{-3}$].}}&&\nonumber
\end{eqnarray}
Using the continuum values $EM_{52} = 68$ and $T_6$ = 15
we find $f \times n_e = 5 \times 10^9$\,cm$^{-3}$. Since the filling factor $f$
can be unity at most, we deduce that $n_{min}=5 \times 10^9$\,cm$^{-3}$ and the
actual density ought to exceed that value. Interestingly, $n_{min}T_{\rm
apex}/T_{\rm O \mathsc{vii}}$ is close to the density determined from demanding
that the high temperature material is at the same pressure as the O\,{\sc vii}
emitting material (assuming that it is immersed in the primary's radiation
field). Alternatively,
density values of $1.5 \times 10^{10}$\,cm$^{-3}$ lead to filling factors
of $\approx$ 0.3. These densities can only be reached without the illuminating
UV radiation field of the primary component. The overall phase range covered
by the {\it Chandra} LETGS observations is 0.32. During that period the
star rotates by 115$^{\circ}$; therefore most of the hemisphere visible at the
beginning of our observations are not visible at the end. The observed
light curve (cf. Fig.~\ref{lightcurve1}) with its slowly decreasing trend does
not appear to give the impression of coming from a number of unrelated regions,
rather, it appears to come from the same region on the star. This is only
possible if the region is located near the pole (in which case it would be
immersed in the radiation field) or if it is fortuitously located in a
equatorial region which happens to be visible during the whole phase interval.
While the latter possibility cannot be excluded, the former option appears more
plausible given the previous evidence for polar activity in Algol
(\cite{schmitt99} 1999), the available VLBI images of Algol (\cite{mutel98}
1998) and the
general occurrence of polar spots in Doppler images in general. It is attractive
to interpret the observed X-ray emission as arising from the circumpolar
regions, possibly from a long-duration flare similar to the one observed
by \cite{schmitt99} (1999). Assuming as height the limit inferred
from the eclipsing BeppoSAX flare (H $\le$ 7 10$^{10}$\,cm) and as filling factor
arbitrarily a value of 1 percent (which is the surface area of the circumpolar 
polar cap), we find a volume of $4\times 10^{32}$\,cm$^{3}$, which requires
densities of $2.5\times 10^{10}$\,cm$^{-3}$ when combined with the observed
emission measures. These densities refer to the hot material at temperatures of
$\approx 15$\,MK. Clearly, these densities are fully consistent with
the observed upper limits. 

\begin{figure}[!ht]
 \resizebox{\hsize}{!}{\includegraphics{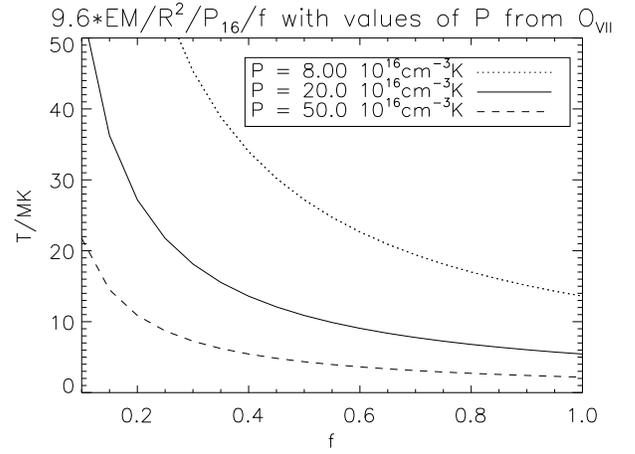}}
 \caption{Temperatures required for asserting certain filling factors
(cf. Eq.~\ref{tf}) assuming constant pressures calculated from the densities and
temperatures measured for O\,{\sc vii}. For the emission measure we use the
value obtained from the continuum $EM=68\times 10^{52}$\,cm$^{-3}$.}
 \label{tf_plot}
\end{figure}

Continuing in our picture of ``canonical'' loops, we can compare a typical
pressure scale height $H_p=5000\, T_{\rm apex} / (g/g_\odot)=3.8\times
10^{11}$\,cm using $T_{\rm apex}=15$\,MK and $g/g_\odot=0.2$,
with a typical loop size $L\approx 30\times 10^8$\,cm (Eq.~\ref{rtv}). Since
$H_p >> L$, the assumption of constant pressure is well justified. Hence
$P = n_e*T$ must be constant and thus Eq.~\ref{feq} can be rewritten as
\begin{eqnarray}
\label{tf}
f=9.6\frac{EM_{52}}{P_{16}R_{11}^2T_6} \Longleftrightarrow
T_6=9.6\frac{EM_{52}}{P_{16}R_{11}^2f}\\
\mbox{\footnotesize[$P_{16}=n_e*T/10^{16}$\,cm$^{-3}$\,K],}\nonumber
\end{eqnarray}
which relates the unknown filling factor with the - in principle - measured
parameters stellar radius, emission measure and coronal pressure.
In Fig.~\ref{tf_plot} we plot the dependence of temperatures on filling factor
for different pressures, using Eq.~\ref{tf}; the emission
measure obtained from the continuum (Sect.~\ref{cont}) was used. The
pressures are calculated from the measured temperatures (Tab.~\ref{ftemp}) and
densities (Tab.~\ref{dens_tab}) for O\,{\sc vii}. In the case of the radiation
field from the B star to influence the density diagnostics, a lower limit for
the pressure of $P=8\times 10^{16}$\,cm$^{-3}$\,K, using the peak formation
temperature $T=2.2$\,MK is calculated, while we obtain $P=50\times
10^{16}$\,cm$^{-3}$\,K for the higher density from omitting the possible effects
from the radiation field and the higher temperature obtained from the
Ly$_\alpha/$r ratio (Tab.~\ref{ftemp}).

It is clear that $T$ must exceed 10\,MK and is very likely below 30\,MK. For low
pressure (P$_{16} = 8$) the filling factor would have to exceed 0.45, while for
high pressure (P$_{16} = 50$) the filling factor would have to be in the
interval $0.07 < f < 0.21$. Since we feel that filling factors $f \sim 1$ are
unlikely, we favor a high pressure scenario.\\

\section{Conclusions}

The high-resolution spectrum obtained with the LETGS on board the Chandra
observatory shows a large number of emission lines and an unusually strong
continuum. We analyze both the emission lines and the continuum relying
exclusively on ratios of individual lines. The LETGS observation interval
covered the orbital phases between 0.74 to 1.06, the light curve appears
to indicate a relative state of quiescence. A slow decline is seen, but cannot
be attributed to the decay of a giant flare, since no cooling and no softening
of the overall emission is detected. The total luminosity is
$L_X=1.4\,10^{31}$\,erg/sec, roughly consistent with the X-ray
luminosities found previously with Einstein and ROSAT.\\

We analyzed the continuum in order to determine an upper temperature and an
overall emission
measure. The continuum can well be modeled with a bremsstrahlung continuum and
a temperature of 15\,MK can be derived which is consistent with the peak
formation temperature of, e.g., Si\,{\sc xiv} which is clearly detected.
Optical depth effects leading to resonant line scattering were analyzed and
ruled out. The ratio of Fe\,{\sc xvii} lines at 15\,\AA\ and 15.27\,\AA\ are
used for testing optical depth effects, but inconsistencies in the atomic line
data were found. From our findings in comparison with other measurements with
the LETGS we conclude that a line ratio of $\approx 2.8$ must occur in plasmas
with $\tau=0$.\\

The measurement of plasma temperatures is carried out using line ratios
of ions from the same element in adjacent ionization stages as, e.g.,
Ly$_{\alpha}$ and He-like resonance lines. At least two temperature components
are found, allowing N\,{\sc vi} (3.2\,MK) and Si\,{\sc xiv} (14.2\,MK) to be
formed. This is consistent with our findings for Fe\,{\sc xxi}/Fe\,{\sc xxii}
($\approx 10\,$MK).\\

For the density diagnostics with the He-like triplets the UV radiation field
originating from the B star companion was analyzed. We found significant effects
for the ions N\,{\sc vi}, O\,{\sc vii}, and even for Ne\,{\sc ix}, however,
the illumination geometry of the primary B-type star is unclear. We therefore
considered both the case with full illumination and no illumination of the
coronal plasma.The determined densities range around
$10^{10}-10^{11}$\,cm$^{-3}$, while for the ions Si\,{\sc xiii} and Ne\,{\sc ix}we find only upper limits. This is also found for all Fe\,{\sc xxi} line ratios,but these upper limits are consistent with densities
$10^{10}-10^{11}$\,cm$^{-3}$ as well.\\

A detailed analysis discussing structural information deduced from the derived
temperatures and densities is presented. The assumptions used for this
discussion are that Algol's corona is composed of a multitude of individual but
identical (semi circular canonical) loops, all of which obey the RTV loop
scaling equation. We further assume that the material is in thermal pressure
equilibrium at 15\,MK. From these assumptions we conclude from our measurements
that these loops have semi-lengths of $L=9\times 10^8$\,cm. The assumption of
the emitting plasma to be immersed in the external radiation field leads us to
lower densities and thus loop lengths of up to $L=460\times 10^8$\,cm, which is,
however, still much smaller than the stellar radius.
The emission measure derived from the analysis of the continuum in combination
with the densities allows us to define constraints on the coronal
filling factor. However, different pressures ranging between 8 and $50\times
10^{16}$\,cm$^{-3}$\,K are possible, depending on whether the emitting plasma is
immersed in the external radiation field or not. A low pressure plasma (i.e.,
effects from the radiation field are severe) requires a large filling factor,
while the high pressure scenario allows a lower filling factor.\\

\begin{acknowledgements}
J.-U.N. acknowledges financial support from Deutsches Zentrum f\"ur Luft- und
Raumfahrt e.V. (DLR) under 50OR98010.\\
The Space Research Organization Netherlands (SRON) is supported financially by NWO.
\end{acknowledgements}


\clearpage

\end{document}